\documentclass{aastex}

\newcommand{\HII}{H\,{\sc ii}}

\shorttitle {Subarcsec-Resolution Radio Maps of Nearby Spirals}
\shortauthors{Tsai et al.}

\begin{document}

\title{Subarcsecond-Resolution Radio Maps of Nearby Spiral Galaxies}

\author{Chao-Wei Tsai\altaffilmark{1}, Jean L. Turner\altaffilmark{1}, Sara C. Beck\altaffilmark{2}, Lucian P. Crosthwaite\altaffilmark{3}, Paul T. P. Ho\altaffilmark{4,5}, \& David S. Meier\altaffilmark{6,7}}

\altaffiltext{1}{Department of Physics and Astronomy, UCLA, Los Angeles, CA 90095-1547; email: cwtsai@astro.ucla.edu, turner@astro.ucla.edu}
\altaffiltext{2}{Department of Physics and Astronomy, Tel Aviv University, Ramat Aviv, Israel; email: sara@wise1.tau.ac.il}
\altaffiltext{3}{Northrop Grumman, San Diego, CA; email: lpcrosthwaite@cox.net}
\altaffiltext{4}{Harvard-Smithsonian Center for Astrophysics, Cambridge MA 02138; email: ho@cfa.harvard.edu}
\altaffiltext{5}{Institute of Astronomy and Astrophysics, Academia Sinica, PO Box 23-141, Taipei 106, Taiwan}
\altaffiltext{6}{Jansky Fellow: National Radio Astronomy Observatory, P. O. Box 0, Socorro, NM. 87801; email: dmeier@nrao.edu}
\altaffiltext{7}{Department of Astronomy, University of Illinois, Urbana, IL}

\begin{abstract}

We report subarcsecond-resolution \textit{VLA} imaging of four nearby spiral galaxies: IC 342, Maffei II, NGC 2903, and NGC 6946. In each galaxy, compact radio continuum sources are identified in the central $\sim$ $15\arcsec \times 15\arcsec$ region. These compact sources are responsible for 20 - 30\,\% of the total emission from the central kpc of the host galaxies at 2 cm, but only $\sim$ 5 - 10\,\% at 6 cm. More than half of the compact sources appear to be \HII\ regions. The \HII\ regions with rising spectra must be fairly dense ($n_{i} \sim 10^{4}~cm^{-3}$) and are presumably very young. The largest of these \HII\ regions require the excitation of 500 - 800 O stars, within regions of only few parsecs extent. These clusters approach the sizes expected for globular clusters. Thermal free-free emission from compact sources contributes more significantly at 2 cm, while diffuse synchrotron emission dominates at 6 cm. The radio \HII\ regions are found near the centers of giant molecular clouds in projection, and do not have obvious visual counterparts.

\end{abstract}


\section{Introduction}

Young massive star clusters, potential young globular clusters, have been revealed by ground-based imaging \citep{1985AJ.....90.1163A} and \textit{Hubble Space Telescope} (\textit{HST}) observations (see review by \citealt{2003dhst.symp..153W}). These clusters appear to be a few Myr in age, $10^{3}$ - $10^{7}~M_{\sun}$ in mass, and 10 pc or smaller in diameter. The youth of these clusters means that, unlike globular clusters, their luminosities are dominated by a population of massive stars. Young, massive star clusters appear to be common \citep{2002AJ....123.1389B} in the centers of late-type spirals, and may play an important role in galactic nuclear activity. How these large star clusters form and the environments favoring their formation are not understood. The identification of young super star clusters in nearby galaxies gives us the chance to study these clusters in the process of formation.

In nearby starburst galaxies, extremely compact sources with rising spectral indices at centimeter wavelengths have been discovered, which are thought to be ``supernebulae" surrounding very young super star clusters (SSC) \citep{1998AJ....116.1212T, 1999IAUS..193..758T, 1999ApJ...527..154K, 2000AJ....120..244B,2000A&A...358...95T, 2002MNRAS.334..912M}. These nebulae have the characteristics of dense, compact \HII\ regions in the Galaxy, and by analogy are also expected to be very young, $\sim$1 Myr or less. However, they are much larger than Galactic compact \HII\ regions, requiring the excitation of hundreds to thousands of young O stars \citep{1998AJ....116.1212T}. These young clusters may be optically obscured by their natal gas and dust clouds \citep{1989ApJ...340..265W}, particularly in molecular gas-rich galactic centers. Free-free emission from embedded \HII\ regions at radio wavelengths is unaffected by extinction from molecular clouds. Radio continuum emission can be used to search for the youngest, potentially embedded young \HII\ regions in other galaxies.

The goal of this investigation is to identify young SSC candidates in four nearby and well-studied spiral galaxies, IC 342, Maffei II, NGC 2903, and NGC 6946 using subarcsecond radio continuum imaging. The centers of these galaxies are infrared-bright \citep{1980ApJ...236..441B, 1980ApJ...235..392T, 1985ApJ...290..108W, 1989ApJ...344..135H} and molecular gas rich \citep{1977ApJ...218L..51R,1982ApJ...258..467Y,1985ApJS...57..261V}. Previous, lower ($\sim 1\arcsec$) resolution \textit{VLA} maps have shown that the four galaxies have concentrated regions of star formation in their centers \citep{1983ApJ...268L..79T, 1994ApJ...421..122T, 1985ApJ...290..108W}. However, the low resolution of these maps, $\sim$ 20 - 70 pc on the galaxy, combines emission from \HII\ regions, supernova remnants (SNRs), and extended non-thermal emission within the beam \citep{1992ARA&A..30..575C}. To isolate nebulae around individual star clusters we need subarcsecond resolution.

Subarcsecond imaging with the \textit{VLA} in its extended configuration maximizes sensitivity to bright and compact radio nebulae over low brightness disk synchrotron emission. Observing at shorter wavelengths, $\lambda<$ 2 cm, also minimizes the contribution of synchrotron emission, which falls with frequency. We present maps of the radio continuum emission at 6 and 2 cm in IC 342, Maffei II, NGC 2903, and NGC 6946 with the A configuration of the \textit{VLA}. The images have resolutions of 0\farcs3 at 6 cm and 0\farcs1 at 2 cm, corresponding to size scales of 3 (IC 342 and Maffei~2) - 11 (NGC 6946) pc on the galaxies.

\section{Observations}

Table 1 lists basic properties for the four spiral galaxies. In all four there is prior evidence for compact radio continuum sources \citep{1981AJ.....86.1175V, 1983AJ.....88..138V, 1983ApJ...268L..79T, 1985ApJ...290..108W, 1994ApJ...421..122T}. The radio data were acquired at the NRAO Very Large Array{\footnote{The National Radio Astronomy Observatory is a facility of the \textit{National Science Foundation} operated under cooperative agreement by Associated Universities, Inc.}} using the A configuration. Previously unpublished 6 cm continuum data were obtained on 1988 November 18 and 20 (program AT98), and 2 cm data on 1983 October 12 (program ID AH141) and 1999 August 23 (program AT227). The 2~cm observations in August 1999 were made using fast-switching, with cycle times of 2 minutes. On source integration times were $\sim$ 1 hour at both wavelengths. Data calibration was done by using \textit{AIPS}, following high frequency reduction procedures for the 2~cm calibration. The sources 3C 48, 3C 138, 3C 147, and 3C 286 are used as flux calibrators. The uncertainty in the absolute flux scale is $\lesssim 5~\%$. Observational parameters are shown in Table 2.

To enhance the sensitivity, we combined our \textit{VLA} A-configuration data with previously published \textit{VLA} archival data listed in Table 2. Data sets of A-, B-, or C-configurations with time on source $>$ 10 minutes and phase center within 18 arcsec (1/10 of \textit{VLA} primary beam at 2~cm) from centers of our measurements were used. Observations with B1950 equinox coordinates were precessed to J2000 coordinates; uncertainty for this conversion is $\sim$20 mas. The uncertainty in the absolute position is determined by the B1950 calibrator positions, which are good to 50 mas. 

Spectral index measurements were done with a separate set of maps with matching (\textit{u,v}) coverages at 6 and 2~cm, which include matched shortest baselines of A-configuration at 2~cm and longest baselines of A-configuration at 6~cm. The largest angular scales sampled by the images are $\sim$ $6\arcsec$. Fluxes and peak fluxes are therefore lower limits to the total flux if extended emission is present. The images were then convolved to the same beamsize, to have matching maximum baseline lengths. Final rms noise levels in blank regions of maps are $\sim$ 0.05, 0.04, 0.03, and 0.03~mJy/beam for IC 342, Maffei 2, NGC 2903, and NGC~6946 respectively at 6~cm, and $\sim$ 0.09, 0.1, 0.05, and 0.14~mJy/beam at 2~cm.

While it is impossible to precisely match (\textit{u,v}) coverages nor to quantify the differences without prior knowledge of the extended source structure, the (\textit{u,v}) plane is well sampled and we estimate that differences between the maps caused by the differences in undersampled extended emission ($>10$\arcsec) must be less than the total resolved out flux divided by this area, which comes to less than our noise level. Since the compact sources we identify are all much less than 10\arcsec\ in extent, the effects on our fluxes with differences in (\textit{u,v}) coverages will be undetectable. 

Interferometer maps are high-pass filtered images and it is difficult to assign uncertainties without knowing the extended structure, which is not known a priori. Point sources, and sources that are well-sampled by the (u,v) coverage will have uncertainties that are described well by the rms noise. Most of our sources fall into this category. We have fitted sizes to the extended sources to better estimate flux uncertainties for these sources.

Although the differences between 6 and 2~cm spatial sampling may not affect compact source fluxes, we stress that there is extended unresolved flux in these maps. One generally compares single dish fluxes to mapped \textit{VLA} fluxes to estimate how much extended emission is resolved out of the \textit{VLA} maps. In this case, the disparity between single dish beams, typically a few arcminutes in size, and our sub-arcsecond beams makes this comparison difficult. We therefore compare our compact source fluxes to lower resolution VLA maps in B and C configuration (shown in Table 1) to see how much flux is resolved out; these lower resolution maps are sensitive to structures up to $45\arcsec$ across at 6~cm and 2~cm, except the case of NGC~2903, in which the C-array flux at 6~cm (sensitive to structure $\gtrsim 120\arcsec$ in size) has been quoted. The total mapped fluxes indicate that our maps recover 15-45\% of the 6~cm flux in the lower resolution observations, and 20-45\% of the 2 cm flux, in both compact and extended components.

\section{The Radio Continuum Images and the Compact Radio Sources}

The 6~cm radio continuum maps and \textit{K}-band (2.2$\mu$m) images of the four galaxies are shown in Figure 1. The naturally-weighted radio contours are overlaid on \textit{2 Micron All Sky Survey} (\textit{2MASS}) \textit{K}-band images on the left. In each of the galaxies the nucleus is a strong $2~\mu$m source and the 6~cm continuum is strongest at the center. We see emission from the central $\sim$ 150 pc except in the case of Maffei II, in which the emission extends over a distance of $\sim$ 350~pc.

The 6~cm maps on the right-hand side of Figure 1, contoured at the same 4~$\sigma$ levels as in the overlay on the left-hand side, are closeups of the radio emission region within the central $16\arcsec$ - $26\arcsec$. In each source there are strong compact sources embedded in extended emission. The western part of the extended emission in IC 342 contains 5 sources aligned from north-west to south-east (named A - E in Table 3) while the eastern part has fewer sources and their emission is also significantly weaker. Ten sources are identified in Maffei II, spread out over 10$\arcsec$ ($\sim$ 240~pc) north-south on the eastern edge of the extended emission. 6 cm continuum emission in NGC 2903 and NGC 6949 is confined to single central sources surrounded by extended emission. The diffuse emission is ``patchy'' because the large scale extended emission has been resolved out by the high resolution. We discuss features in the individual galaxies in subsequent sections. Peak flux densities at 6~cm are 1.2, 2.6, 0.4, and 1.9~mJy/beam for IC 342, Maffei II, NGC 2903, and NGC 6946, respectively.

The radio maps are overlaid on \textit{HST} \textit{V}-band and H$\alpha$ images in Figures 2--5. For convenience in comparing to the 2~cm, the 6~cm maps in Figures 2 - 5 were restricted to the same (\textit{u,v}) coverage ($\sim$ 30 - 600~$k\lambda$), and convolved to the same beamsize. The radio and the HST images have similar resolutions. The uncertainty in the registration of the images is $\sim$1\arcsec, limited by the astrometry of the HST images. In IC 342, Maffei 2, and NGC 2903 the radio sources tend to lie in regions of apparent dust obscuration. In NGC 6946 the extended central radio continuum source nearly aligns with a bright optical source at the center. The uncertainty in registration does not allow us to determine how precisely they align.

In Table 3, we list compact sources detected in the central 15$\arcsec$ radius of each galaxy and their fluxes at 6 cm and 2~cm, with proper (\textit{u,v}) restrictions described in $\S2$. The sources meet at least one of the following criteria: (1) $5~\sigma$ detection at the peak intensity at one wavelength, or (2) $4~\sigma$ emission detection in both wavebands. The sources are named alphabetically following radio convention in order of their peak flux densities at 6 cm. Positions and deconvolved sizes of sources were obtained from the 2~cm data fit assuming a single Gaussian source using the \textit{IMFIT} task in \textit{AIPS}. For sources absent at 2 cm, Gaussians were fit to the 6~cm maps. The uncertainty in size is $\lesssim$ 30\,\% of the beam. The total integrated flux densities were measured using the \textit{AIPS} task \textit{IRING}. The upper limits quoted for non-detections are $3~\sigma$. The systematic error of flux is expected to be $\lesssim$ 5\,\%, related to the uncertainties in the absolute flux calibration fluxes. We separated this uncertainty from signal to noise, which is listed in the Table 3.

In Table 4, we list the properties of each compact radio source. The spectral index is defined as $S_{\nu} \propto \nu^{\alpha}$. The uncertainties of $\alpha_{6-2}$, the spectral index between 2 and 6~cm wavelengths, are derived with $\sqrt{2}~\sigma$ variances of flux densities in both wavebands; uncertainties are typically $\pm$0.2-0.3. For sources A - D and I in the extended emission complex of NGC~6946, we derive the $\alpha$ from their intensities, which are well-determined, rather than their fluxes, which are more uncertain. Based on the values of $\alpha_{6-2}$ (discussed in the next section), we classified sources as optically thick \HII\ regions (\HII, thick), optically thin \HII\ regions (HII, thin), supernova remnants (SNR), or radio supernovae (RSN). Source types are listed in the fourth and fifth columns in Table 4. For the cases identified as \HII\ regions, $N_{Lyc}$ in the sixth column is the Lyman continuum rate derived from the flux density at 2 cm on the assumption of optically thin emission and electron temperature in \HII\ region, $T_{e} \sim$ 10,000~K. If the emission is optically thick, this value will be a lower limit to the true $N_{Lyc}$. Electron temperatures of \HII\ regions tend to be cooler toward the high-metallicity centers of spiral galaxies \citep{1979MNRAS.189...95P, 1981PNAS...78.1994S}. The values of $N_{Lyc}$ will be $\sim$ 15\,\% lower than listed, if $T_{e} =$ 6,000~K rather than the adopted $10^4$~K. $N_{O7}$ is the number of standard O7 stars needed to generate $N_{Lyc}$. We also assume ionization-bonded nebulae; if UV photons escape the \HII\ region, our OB luminosities will be underestimates of the true Lyman continuum rates.

\section{Discussion}

\subsection{Identification of Luminous Compact \HII\ Regions from their Radio Spectra}

The radio images reveal a population of bright compact radio sources in the centers of IC 342, Maffei 2, NGC 6946, and NGC 2903. The sources are comparable in extent to the 0\farcs3 beam, or $\sim$ 5 - 10 pc in diameter. What are these sources?

The cm-wave continuum emission from nearby normal galaxies is a mixture of synchrotron emission from SNRs and disk cosmic rays, and thermal bremsstrahlung emission from \HII\ regions.  Spectral index $\alpha$ is an indicator of the emission mechanism: optically thin free-free emission has $\alpha \sim -0.1$, while optically thin non-thermal emission by synchrotron radiation will have steeper negative spectral indices. Galactic SNR have $\alpha_{nt} \sim$ $-0.6$ to $-0.2$, with the majority in the range $-0.5$ to $-0.4$ \citep{1984MNRAS.209..449G, 1984AuJPh..37..321M, 1986ApJ...301..790W}. Steeper $\alpha_{6-2} < -1.0$ are characteristic of short-lived RSNe \citep{1986ApJ...301..790W}, and of extended synchrotron disk emission.

\HII\ regions are optically thick at wavelengths longer than a turnover frequency that depends on the electron temperature and emission measure, EM=$\int n_e^2~dl$. At long wavelengths, \HII\ regions are bright, and their fluxes rise with frequency until the turnover frequency. The higher the emission measure, the higher the turnover frequency. Spectra of diffuse or ``classical" ($\rm EM\le 10^4~cm^{-6}~pc$) \HII\ regions that appear in visible wavelengths typically turn over at meter wavelengths and are optically thin at cm wavelengths. A cm-wave turnover frequency is characteristic of compact or ultracompact \HII\ regions \citep{1967ApJ...147..471M}, which have $\rm EM \ge 10^8-10^9~cm^{-6}~pc$. 

Beyond the Local Group, ultracompact \HII\ regions typical of those in our Galactic disk, excited by individual O stars or small clusters, are small in size \citep{1979ARA&A..17..345H} and thus too faint to be seen \citep{1994ApJ...421..122T}. The compact component \citep{1967ApJ...150..807M} of the luminous Galactic \HII\ region W49 would fail to meet our detection criteria in our closest galaxy (IC 342), as would the radio source \citep{1999AJ....117.2902D} of NGC 3606. However, if the clusters are somewhat larger, with at least 200 O stars surrounded by an ionization-bounded volume of gas, even dense, compact \HII\ regions, with $\rm EM \ge 10^8-10^9~cm^{-6}~pc$, can be detected in external galaxies \citep{1998AJ....116.1212T}. Even for the largest clusters ($N_{O7} = 10^{4}$), dense, compact \HII\ regions will still be small, less than 2 - 3 pc in size. Since large numbers of massive stars can also be accompanied by supernovae, thermal free-free emission from these \HII\ regions can be mixed with nonthermal emission within a single beam if the resolution is low \citep{1998AJ....116.1212T, 2000ApJ...532L.109T}. Confusion is minimized if the resolution is high, and the galaxy is nearby, so that the beam closely matches the size of a single \HII\ region. Subarcsecond imaging maximizes our sensitivity to these sources. 

More than half of the 38 identified compact sources of Table~4 appear to be \HII\ regions. A significant fraction may have rising spectra, indicating that they are at least partly optically thick at 6~cm. Given the likely presence of nonthermal synchrotron emission in the vicinity of these star-forming regions, the spectral indices could systematically be even higher than we measure here, due to spectral confusion \citep{2000ApJ...532L.109T}. Rising cm-wave spectra imply that the rms electron densities are $\sim 10^3-10^{4}~cm^{-3}$. That a large percentage of our sources are ultracompact \HII\ regions might be expected from the subarcsecond resolution, which picks out these high brightness regions. There may be lower density, extended and less bright \HII\ regions that are below the brightness limit of these maps.

\subsection{Compact Nebulae as Young Star Formation Tracers}

\subsubsection{Properties of the Compact \HII\ Regions}

The sizes of the compact \HII\ regions listed in Table 3 are obtained from deconvolution of the beam from the observed source in the 2 cm map, assuming that the sources are gaussian. (They are probably not gaussian in structure, but this gives us approximate source sizes.) Uncertainties in the sizes, due to beam size, source structure, and signal-to-noise ratio, are $\lesssim$ 30\,\% of the beam. The derived  source sizes in IC 342, Maffei II, and NGC 2903 are $\sim 0.3\arcsec$, but $\sim$0.5$\arcsec$ in NGC 6946. Sources with larger deconvolved sizes in NGC 6946 are nebulae joined with the large emission complex at the center, and suffer from confusion. 
Seeing may be a problem at 2~cm. The similarity of the source sizes suggest this may be the case;  it is likely, therefore, that the sizes we obtain are upper limits to the true source sizes. 

The deconvolved sizes  correspond to diameters of $\sim$ 6 - 14~pc. These \HII\ regions are significantly smaller than the 30 Doradus \HII\ region ($\sim$ 200~pc in diameter for region of $EM > 10^{4}~pc~cm^{-6}$; see \citealt{1978MNRAS.185..263M} and \citealt{1984ApJ...287..116K}) in the Large Magellanic Cloud (LMC). They are larger but similar in nature to the Galactic compact \HII\ regions in W49 \citep{1967ApJ...150..807M,2002ApJ...564..827C} and NGC 3603 \citep{1999AJ....117.2902D, 2004AJ....127.1014S, 2002ApJ...571..366M}. One possibility for the fact that these \HII\ regions are relatively compact and dense is that they are younger than 30 Doradus or NGC 3603. Because these regions are 
nearly unresolved, and because we do not have good radio SEDs, at this stage we cannot model their properties with precision.

The estimated sizes of these nebulae are similar to the pc-sized diameters of the supernebulae found in NGC 5253 and II Zw 40 \citep{2004ApJ...602L..85T, 2002AJ....124.2516B, 2002AJ....124..877M}. These \HII\ regions have ionized gas densities at least as high as $n_{i} \sim 10^{3} - 10^{5}~ cm^{-3}$; they are dense, with high pressures. The compactness of these optically thick \HII\ regions implies some degree of youth, although \HII\ regions around massive young clusters may differ dynamically and evolve differently from Galactic compact \HII\ regions \citep{2002MNRAS.336.1188K, 2003Natur.423..621T, 2004ApJ...602L..85T}.

The radio fluxes of these compact \HII\ regions require that they harbor hundreds of massive stars. The luminosities of the clusters exciting the \HII\ regions can be inferred from their Lyman continuum rates using $\rm L_{OB}=2\times10^{-44}~N_{Lyc}$ for a Salpeter IMF. The inferred luminosities are not very sensitive to IMF since luminosity, like ionizing photons, is dominated by the most massive stars. The cluster luminosities inferred from the free-free emission are $\sim 10^8-10^9~L_\odot$. Based on their luminosities, the largest of these radio clusters have roughly an order of magnitude more stars than the R 136 cluster within 30 Doradus \citep{1998ApJ...493..180M, 2002IAUS..207..531Z}, or in the clusters within the large Galactic \HII\ region NGC 3603 \citep{1998ApJ...498..278E,1999AJ....117.2902D} or W49 \citep{2002ApJ...564..827C}.

The total Lyman continuum rate, $N_{Lyc}$, is derived from the total flux from thermal compact sources, assuming the emission is optically thin and the \HII\ regions are ionization-bounded. These values are compared with $N_{Lyc}$ rates derived from millimeter continuum, which are unlikley to suffer from opacity effects. In IC 342, the estimated $N_{Lyc}$ is $\sim 6 \times 10^{51}~s^{-1}$ for six thermal compact sources. This can be compared with the value of $N_{Lyc} \sim 3 \times 10^{52}~s^{-1}$ obtained by \citeauthor{2001ApJ...551..687M} (\citeyear{2001ApJ...551..687M}; corrected for distance) for the central $16\arcsec$ region derived from 2.6 mm continuum, and $N_{Lyc} \sim 1.2 \times 10^{52}$ for the two major 2.6 cm peaks corresponding to the central starburst regions. Over the central $7\arcsec \times 2\arcsec$ region of NGC 6946, $N_{Lyc}$ is $3 \pm 0.2 \times 10^{52}~s^{-1}$, as derived from the 2.7 mm images of \citeauthor{2004AJ....127.2069M} (\citeyear{2004AJ....127.2069M}; corrected to our distance).  Similar agreement is also found in Maffei II between the $N_{Lyc}$ inferred by thermal fluxes in mm \citep{2006.PREP.MEIER} and cm (this work). These numbers imply that the fluxes of compact sources make up at least 50 \% of the total $N_{Lyc}$ photons seen towards the starburst peaks in lower resolution images. The fact that the mm emission gives higher values for $N_{Lyc}$ suggests that some of the thermal sources are optically thick at cm wavelengths, or that some of the thermal flux is extended, or both.

\subsubsection{Compact Nebulae and Molecular Clouds; Possible Star Formation Triggers}

The existence of massive stars within these clusters indicates their youth. Because the local star formation is expected to be suppressed by massive star feedback of mechanical energy through winds and supernovae, we expect the stars in these large clusters to be coeval to within a few Myr. Thus they are presumably located inside and co-moving with their parent molecular clouds, and should be close to the sites at which they formed. Given that these \HII\ regions are compact and quite dense, it is likely that they may be significantly younger than the $\sim$ 10 Myr we might infer from the presence of massive stars.

The young, massive clusters in IC 342 and Maffei II (shown in Figure 2 and 3) are aligned with dust lanes and molecular gas. These nebulae are found in coherent structures extending over regions 110 pc (IC 342) - 240 pc (Maffei 2) in size. The sound crossing times for these distances are 10 - 25~Myr, if the sound speed is $\sim$10~km/s. This timescale is comparable to or longer than the lifetimes of massive stars. From the youth inferred from the high gas densities and from the spatial distribution of these \HII\ regions along the CO arms, we conclude that the formation of young, massive stellar clusters in compact nebulae found in IC 342 and Maffei II is mainly triggered by large scale resonant features \citep[i.e., gas arms associated with spiral density wave/bar features;][]{1984ApJ...282L..59L, 1990Natur.344..224I, 1989ApJ...344..763I, 1991ApJ...377..434H, 2003ApJ...591L.115S, 2004AAS...205.9909M, 2005ApJ...618..259M}. This is consistent with their locations with respect to optical dust lanes (Figures 2-5).

Comparing with the \textit{HST} images (Figure 2 - 5), we find that only a small fraction of the radio \HII\ regions have possible optical counterparts. The extinction in embedded \HII\ regions is often internal to the nebulae \citep{1986ApJ...309...70B, 1989ApJ...337..230K, 1990ApJ...349...57H}, but it is also possible that the compact nebulae are hidden behind the foreground clouds or embedded in their host giant molecular clouds. If we assume that all of the luminosity comes out in the infrared, the implied $\rm L_{IR}$ of our compact sources amounts to only 3 - 5\,\% of the total IR luminosities of the galaxies (Table 1).

The young nebulae are all located within the central $\sim$ 100~pc except in Maffei II, where the \HII\ regions extend over $\sim$ 250 pc. To support the formation of numerous massive stars, there should be high molecular gas concentrations \citep{1991ARA&A..29..581Y}. In all four  galaxies, the central regions have molecular gas surface densities 3 - 9 times higher \citep{1991ApJ...377..434H, 1992ApJ...384...72T, 2001ApJ...551..687M, 2004AJ....127.2069M} than do typical spiral galaxies on size scales of $6\arcsec$ \citep{2003ApJS..145..259H}. The high concentrations of molecular gas ($\Sigma_{gas} = 600 - 2000~M_{\sun}~pc^{-2}$ within central $6\arcsec$; \citealt{1991ApJ...377..434H, 2003ApJS..145..259H}) lead to high internal visual extinctions $A_{V}$ of up to tens to hundreds of magnitudes \citep{1990ApJ...349...57H, 1991ApJ...377..434H, 1992ApJ...384...72T} toward these galactic centers.

The young star clusters associated with the radio nebulae are preferentially found projected onto the centers of giant molecular clouds (\S 4.3). This means that there is a wealth of gas around the star clusters to fuel more star formation, unlike the high star formation efficiency inferred from the lack of gas in the super star clusters found in the dwarf galaxy NGC 5253 \citep{1997ApJ...474L..11T, 2002AJ....124..877M, 2002AAS...20111616T}. Given the proximity of molecular clouds to these clusters, and their location in the centers of large spiral galaxies, it is unlikely that any of the young clusters associated with the radio nebulae in these four spirals will become long-lived, gravitationally bound clusters, i.e., globular clusters.

\subsection{Upper Part of the Luminosity Function of Young \HII\ Regions}

Nearly two dozen radio sources in the four galaxies were found to have flat or rising spectra (Table 4). These are presumably thermal nebulae excited by massive stars, for which the 2 cm fluxes give a lower limit to the required $N_{lyc}$ ($\S$ 4.1). In Figure 6, we show the Lyman continuum luminosity function (LF) of these \HII\ regions. The corresponding $H_{\alpha}$ luminosities have been calculated from the $N_{Lyc}$ assuming no extinction, for comparison with other studies. The sample is small, and the statistical error bars are relatively large. However there are some trends that appear to be real and should be studied with a larger sample.

We have fit the LF with a power law, and also with a two component power law. The two component fit is significantly better than the single component fit. The LF has a steeper slope for nebulae with $H_{\alpha}$ luminosity $>2 \times 10^{39}$ erg sec$^{-1}$ (corresponding to $\rm L_{OB}=10^{7.5}~L_{\sun}$, about 150 O7 stars). The solid line in Figure 6 represents power law models ($N(L) \propto L^{a}dL$) of LF with index of $-2.2$. The power law index we obtain is close to the average value of $-2.0$ in 30 nearby spiral and irregular galaxies from an optical study of \HII\ regions by \citet{1989ApJ...337..761K}. However, the slope of the LF for the more luminous \HII\ regions of our sample is shallower than the slope in the two component power law LF, or ``type II'' LF, found in optical studies of other spiral galaxies, M 51, NGC 3521, NGC 3627, NGC 4736 \citep{1989ApJ...337..761K}, and NGC 7714 \citep{1995ApJ...439..604G}. The optical studies also show the turnover point at lower luminosity (at $L_{H_{\alpha}} \sim 10^{38.7} - 10^{39}$). A possible explanation for this difference might be that the luminous sources we found are very young and obscured, thus not detected in the optical, although this may also be small-number statistics.

The flattening of the LF at the faint end may be caused by incompleteness due to both limited sensitivity and partially resolving out extended \HII\ regions. We cannot detect unresolved (diameter $<$ 15 pc) nebulae with $N_{Lyc} < 4\times 10^{50}~ s^{-1}$ (3$\sigma$), corresponding to 40 O stars in IC 342 and 120 stars in NGC 6946. However, we also cannot detect more luminous, but lower
brightness regions if they are large and resolved (size $\gg$ 15~pc); in this limit we are sensitive to \HII\ regions with $EM > 10^4 \rm ~cm^{-6}\,pc$. 

We did the same thermal source identification and $H_{\alpha}$ luminosity calculation for radio sources in M 82, another nearby starburst galaxy, based on radio studies of \citet{1999PhDT........10A}. Allen's source list is more complete, with $N_{Lyc} < 2\times 10^{50}~ s^{-1}$ (20 O7 stars) and radio nebulae with size $\lesssim$ 50 pc. The LF of the 21 sources in M82 with flat or rising spectra between 2 and 6 cm shows very similar features in slope and the break point in LF to what we find in the four spirals \citep{2006EAYAM...T}. The agreement of LF shape of M 82 and our sample suggests that the broken power law may be real. Larger samples would help to clarify this issue.

The steep power law LF suggests a cutoff at the high luminosity end in the young clusters associated with radio nebulae in spiral galaxies. The upper end of the luminosity function is around $L_{H_{\alpha}} \sim 10^{40}~ erg~ s^{-1}$. This corresponds to $\sim$ 750 O7 stars. We observe two such clusters, in Maffei 2 and NGC 2903, and their nebulae are $\lesssim5$ pc in extent. The total mass of such a cluster would be $\sim$ 2.2 $\times 10^{5}~ M_{\sun}$ (assuming Salpeter IMF), consistent with the masses of Galactic globular clusters. 

\section{Individual Galaxies}

\subsection{IC 342}

IC 342 is a face-on giant Scd spiral \citep{1971USNO.2.20.79, 2000AJ....119.1720C, 2001AJ....122..797C} close to the Galactic plane. Its distance is uncertain: \citet{2002AJ....124..839S} used the $P$ - $L$ relation for Cepheids and suggest 3.3 Mpc, close to the 3.9 Mpc given by \citet{1988ang..book.....T}, but earlier measurements found $\sim$ 2 Mpc in optical spectroscopic and photometric studies \citep{1989AJ.....97.1341M, 1993A&AS..100..227K}. Here we adopt $3.3$ Mpc ($1\arcsec = 16$ pc).

The core of IC 342 contains a central molecular ring 5$\arcsec$ - 6$\arcsec$ in radius and a nuclear CO disk $\sim 3.5\arcsec$ ($\sim 55$ pc) in diameter \citep{2003ApJ...591L.115S}. The 5$\arcsec$ - 6$\arcsec$ nuclear disk contains $\sim 2 \times 10^{5} ~ M_{\sun}$ molecular gas and the central 100~pc contains $\sim 5 \times 10^{6} ~ M_{\sun}$ of molecular gas \citep{2001ApJ...551..687M}. The nuclear star cluster (NSC) of IC 342 has been studied in optical and infrared wavelengths using \textit{HST} and \textit{IRTF}. The NSC of IC 342 has mass $M \approx 6 \times 10^{6}~ M_{\sun}$ and age 10$^{6.8-7.8}$ years. The upper limit on a central black hole is $5 \times 10^{5}~ M_{\sun}$ \citep{1999AJ....118..831B}. Outside the central star cluster, there is current star formation indicated by bright IR and radio emission clumps, each containing $10^{2}$ - $10^{3}$ OB stars, and the total number of equivalent O stars is about 5,000 \citep{1980ApJ...236..441B, 1983ApJ...268L..79T}. Our measurements resolve out most of the extended component of emission so our calculated total O star population, which detects compact sources only, is over an order of magnitude less.

The radio continuum images (Figure 2) reveal twelve compact radio continuum sources on the two spiral arms and the central molecular ring of IC 342. 10\,\% of the total radio emission at 6 cm and  20\,\% at 2 cm for the central kpc arises in compact sources. Inferring the infrared luminosities from their radio fluxes and $N_{Lyc}$ (``$L_{OB}$"), we find that the compact \HII\ regions in the nucleus contribute only $\sim$ 5\,\% of the total infrared luminosity of IC 342. 2 cm sources A, C, E, G, and I - K are distributed along the inner edge of the dust lane and the western molecular spiral arm near the nucleus \citep{2003ApJ...591L.115S}. The western arm of the central molecular ring clearly has more energetic star formation than the eastern arm, as is evident from lower resolution observations. The star formation in IC 342 appears to be spatially distributed in the nuclear region, although somewhat concentrated to the west near the northern end of the western molecular arm \citep{2005ApJ...618..259M}.

We identify a number of likely SNR. Source G in the eastern central molecular arm has $\alpha \sim -0.3.$ Four sources, B, D, F, and H, are detected only at 6 cm. Sources A and B have $\alpha \sim -0.5$ and $-0.6$ respectively, close to the conventional $\alpha$ of Galactic SNR. The steep spectrum of Source D makes it a candidate for a RSN, which typically decline rapidly over the span of a few years. However, no obvious change in the 6 cm flux of source D is found in the data of IC 342 we have collected over multiple epochs separated over 15 years. Its diameter ($\sim 0.6\arcsec - 0.4\arcsec$, or 10 pc to 6 pc) and luminosity $L_{6 cm} = 2.9 \times 10^{19} erg \; sec^{-1} \; Hz^{-1}$ can be explained better by a SNR. The unusual negative spectral index of source D might be due to the flux contamination by nearby non-thermal source F. Sources F and H show no strong emission at 2 cm, and have upper limits to their spectral indices of $\alpha < -0.5$.  None of these sources, A, B, D, or F - H, are associated with obvious visible star clusters. As SNR candidates, they are the likely end stages of massive stars near the young star cluster systems.

We identify five compact sources that appear to be compact \HII\ regions. Sources C, I, J, and K in the north-west and source L in the east have spectra consistent with optically thick free-free emission. If the emission were optically thin, the excitation of these compact nebulae requires clusters of at least 70 - 210 O7 stars each. These are lower limits to the true cluster population, because the emission is optically thick, and because of the possibility of dust absorption of the UV photons. The inferred cluster masses are then $2.1$ - $6.3 \times 10^{4}~M_{\sun}$. These are large clusters comparable to or a few times larger than 30 Doradus in the LMC, but much more compact. These nebulae are less than 8 pc in size as compared to 200 pc for 30 Doradus, and hence probably younger. These star clusters are not detected in the \textit{HST} image, implying that $A_{V} \gtrsim$ 5 mag, or $N_{H} \gtrsim 1.0 \times 10^{22}~cm^{-2}$, if the ZAMS star clusters with Salpeter IMF are assumed. The $4\arcsec$ resolution $10~\micron$ image of IC 342 \citep{1980ApJ...236..441B} shows two peaks at ($\sim$~03$^{h}$46$^{m}$48.3$^{s}$; +68$\arcdeg$05$\arcmin$47$\arcsec$) and ($\sim$~47.6$^{s}$; 46$\arcsec$), close to the radio \HII\ regions we identify here. The visual extinction estimated from silicate absorption feature is $\sim$ 15~mag toward these two $10~\micron$ peaks \citep{1980ApJ...236..441B}, consistent with our argument of high extinction toward these \HII\ regions.

To the west, Source E has a flatter radio spectrum and appears to be an optically thin source. This \HII\ region requires about 70 O stars, comparable to 30 Doradus. A ZAMS star cluster with 70 O stars in IC 342 should have apparent brightness $m_{V} = 13.8$ when Galactic extinction of $A_{V} = 1.85$ ($N_{H} \sim 3.5 \times 10^{21}~cm^{-2}$) toward IC 342 is applied; these clusters do not appear in the \textit{HST} optical image (Fig 3, also see \citealt{1999AJ....118..831B}).

We do not detect significant radio continuum emission at small size scales from the nuclear star cluster to a $3~\sigma$ limit of $6.5 \times 10^{50} sec^{-1}$, or 70 equivalent O7 stars. This is consistent with a somewhat older stellar population there \citep[60 Myr][]{1999AJ....118..831B}. No significant active galactic nucleus has been detected with these high resolution measurements, but our own Galactic center at a similar resolution \citep[][]{1983ApJ...268L..85B} would be below our detection limit at this distance ($\sim$ 0.1~mJy at 2 cm).

\subsection{Maffei II}

Maffei II, discovered in the infrared by \citet{1968PASP...80..618M}, is a spiral galaxy with a strong and long bar evident in near-infrared images \citep{1993AJ....105..121H}. It is located behind the Galactic plane ($l=136.50\arcdeg$, $b=-0.33\arcdeg$) at $A_V\sim$ 5 - 10 mag of Galactic extinction. The nucleus of the galaxy is actively forming stars \citep{1983ApJ...268L...7R, 1990ApJ...349...57H, 1993ApJ...414..120T, 1976A&A....48..413S, 1994ApJ...421..122T}. The distance of Maffei II was recently set at 2.8 Mpc by \citet{2003A&A...408..111K} using the Tully-Fisher relation; however the HI linewidth suffers from Galactic absorption, particularly in lower resolution spectra \citep{1996ApJ...466..135H} so the actual H\,{\sc i} linewidth is larger than what they adopt. In this paper, we use the distance of 5 Mpc estimated from the diameters of the largest \HII\ regions in Maffei II \citep{1973ApJ...180..351S} and by the Tully-Fisher analysis of \citet{1993ApJ...404..602H}. The uncertainty in the distance to Maffei 2 introduces a factor of 3 uncertainty in estimates of massive star numbers and total masses, and a factor of 2 uncertainty in source sizes.

In Figure 3 we show the radio continuum maps of Maffei II. Ten sources are identified in both 2 and 6~cm maps and sources B and J are identified only at 6~cm (see Table 3 and Fig 3). The compact sources lie on a twisted, inverse ``S'' shaped line north-south, following the trend of larger scale gas distribution and coinciding with the dust lane shown in the HST \textit{V}-band image. The radio continuum emission is located along the elongated bar of $45\arcsec$ length seen in $^{12}$CO \citep{1989ApJ...344..763I} and $^{13}$CO \citep{1991ApJ...377..434H} images. Northern sources B, C, and F are associated with giant molecular cloud (GMC) D1 of \citet{2006.PREP.MEIER}. Sources A, D, H, I, and J are close to the center of GMC E. The southern sources, E and G, are between GMC E and F. Strong detections in both 6 and 2~cm provide good spectral indices and sizes for the radio sources in Maffei II, except for source B and J which have no significant detection at 2~cm.

The total flux densities from compact sources constitute $\sim$15\,\% of the total emission of the central kpc of Maffei II at 6 cm and $\sim$30\,\% at 2~cm (for structures $< 45\arcsec$; see $\S$2). \citet{1994ApJ...421..122T}, using the \textit{VLA} in B array at 6~cm and C array at 2~cm, found a total number of O stars in good agreement with our results. The star formation in the center of Maffei 2 is concentrated in large, compact clusters aligned in a linear structure along the northern CO arm. The predicted infrared luminosity $L_{IR} = L_{OB} \sim 5.3 \times 10^{8}~ L_{\sun}$ from these point sources is $\sim$5\,\% of the total $L_{IR}$ of Maffei II. Except for three SNR/RSN candidates, the compact radio continuum sources we detect are thermal free-free emission from \HII\ regions. Most of the thermal sources in Maffei II have relatively flat spectra, compared with rising spectrum sources in NGC 2903 and NGC 6946.

Spectral indices $\alpha \sim -0.1$ for sources A, D and I suggest that these are star clusters with optically thin thermal free-free radiation. All three are in the center of the radio emission near GMC D1. These \HII\ regions of diameter $\lesssim$ 10~pc each require $\sim$ 200 - 600 exciting O stars. The inferred masses of the clusters associated with sources C and G are $\sim 6$ - $18 \times 10^{4}~ M_{\sun}$, slightly larger than the young massive star clusters in our Galactic Center.

Sources C, E, G, and H have rising spectra indicating that they are compact \HII\ regions. These \HII\ regions are brighter and more luminous than the ones in IC 342. The lower limit of $N_{Lyc} = 2.9$ - $4.9 \times 10^{51}~s^{-1}$ requires clusters of more than 300 - 500 O7 stars. The inferred cluster masses are $M_{C} \sim 1$ - $1.5 \times 10^{5}~M_{\sun}$. The mass of source C, $M_{C} \sim 1.5 \times 10^{5}~M_{\sun}$ is only a small fraction of the $26 \times 10^{6}~M_{\sun}$ upper limit mass of GMC C \citep{2006.PREP.MEIER}. The star formation efficiency therefore appears to be of order one percent, which is similar to the Galactic value on these scales. From the C$^{18}$O(2-1) map of \citet{2006.PREP.MEIER}, Source E and G seem to be at the gas bridge in the middle of GMC D1 and GMC E3. The predicted visual brightness of these rising spectrum sources with the assumption of a Salpeter IMF is $m_{V} \sim 25$ mag with Galactic extinction, so these sources are too faint and highly obscured to be detected via \textit{HST} observations.

Source F is probably a SNR, with $\alpha  -0.4.$ The luminosity of source F at 6 cm is $L_{6 cm} \sim 4.6 \times 10^{25}~ erg~ sec^{-1}~ Hz^{-1}$ and the deconvolved size is 11 $\times$ 8~pc, characteristic of a young Type II SNR. Sources B and J show no significant counterpart at 2 cm, and their negative spectral indexes suggest that they are RSN and SNR, respectively. All three non-thermal sources, F, B, and J, are within $\gtrsim 1 \arcsec$ (25 pc in projection) to the nearby optically thick thermal sources, C and H. The locations of these SNR/RSN candidates suggest they might be associated with the compact \HII\ regions.

\subsection{NGC 2903}

NGC 2903 is a large and bright ``hot spot'' spiral, with an unusually complex structure of both early- and late-type stars and \HII\ regions at its center \citep{1974PASJ...26..289O, 1967PASP...79..152S, 1958PASP...70..364M, 1973PASP...85..103S}. It is at a distance of 8.9 Mpc \citep{2000A&AS..142..425D}. \citet{1985ApJ...290..108W} studied the ``hot-spot'' galaxy in the radio and infrared, and found that its radio emission is predominantly nonthermal. However,  in some regions the free-free emission contributes substantially at 2 cm. NGC 2903 has a well-studied star-forming ring. IR maps find \HII\ regions in the ring with luminosities approaching that of 30 Doradus \citep{2001MNRAS.322..757A}. IR spectroscopy gives an age range of 4 - 7 $\times 10^{6}$ yr for the embedded star clusters while the \HII\ regions are thought to be only 1 - 4 $\times 10^{6}$ yr of age \citep{2001MNRAS.322..757A}.
 Molecular gas in NGC 2903 is concentrated in the central few tens of arcseconds \citep{2002AJ....124.2581S, 2003ApJS..145..259H}. 

In Figure 4 we show NGC 2903 in radio continuum maps and \textit{HST} \textit{WFPC2} images (\textit{V}-band and H$\alpha$). About 20\,\% of the total flux density at 2~cm emission in the central kpc comes from compact sources B, C, E, F, and G. At 6~cm $\sim$5\,\% is due to sources A - E. This agrees with \citet{1985ApJ...290..108W} on the contribution of compact sources at radio wavelengths. The northern peak near 09$^{h}$32$^{m}$09.99$^{s}$, 21$\arcdeg$30$\arcmin$12.7$\arcsec$ seen in low resolution 2~cm maps \citep{1985ApJ...290..108W} does not have a corresponding compact source in our images, and therefore must be extended over $>$2$\arcsec$ scales.

\textit{K} and \textit{H}-band \textit{NICMOS} and Pa$\,\alpha$ images of the nucleus of NGC 2903 \citep{2001MNRAS.322..757A} show an absence of ionized gas, consistent with what is seen in our 2~cm and 6~cm maps. The nucleus must consist of a relatively old stellar population dominated by giants and supergiant stars. The northern peak of the lower resolution 2~cm maps of \citet{1985ApJ...290..108W} corresponds to compact Source B. It has a rising spectrum of $\alpha \sim 0.8$ through cm wavelengths. Excitation of Source B requires at least 800 O7 stars if the free-free emission is thin and more if thick. This result is $\sim$ 2.6 times higher than $N_{Lyc}$ derived by \citet{2001MNRAS.322..757A}. It suggests visual extinctions of 2.7~mag higher than what \citet{2001MNRAS.322..757A} obtained, thus $A_{V} = 6.5$, assumed if $A_{V}/A_{1.87\micron}$ applied \citep{1989ApJ...345..245C}. This difference could be explained if there is significant extinction at Pa$\,\alpha$. The brightness temperature of Source B indicates a size of $0.025\arcsec$, or 1.1~pc for a nebula with $T_e=$10,000 K. Source B is a good candidate for a super star cluster nebula.

Sources A, D, E, and G are within the extended emission east of the northern peak \citep{1985ApJ...290..108W} near Source B. All of them have counterparts in Pa$\alpha$ \citep{2001MNRAS.322..757A} except Source E. Source E is close to source ``b'' of \citet{1974PASJ...26..289O}, which they claim to be a free-free emitter. Source E also appears coincident with cluster 2 in the \textit{HST} near-infrared images \citet{2001MNRAS.322..757A} and the AO \textit{K'}-band image \citep{2000MNRAS.317..234P}. Sources A, B, D, and G correspond to sources H1, H4, H2, and H3 of \citet{2000MNRAS.317..234P}, respectively. In our maps, Source G has been identified only at 2~cm, and shows a very steep rising spectrum. Its spectral index, $\alpha > 1.0,$ implies that this is a possible compact \HII\ region with a high electron density and should therefore be very young. Source D, detected at 6~cm but not at 2~cm,is probably a SNR. Source A has a flat slope of $\alpha \sim 0.0^{+0.3}_{-0.3}$. It could be a small \HII\ region, with $\sim$ 210 equivalent O stars, or, within the uncertainties, it could also be a single young Crab-like SNR with the luminosity of Cas A.

Southern Sources C and F have thermal spectra. Source C coincides with the southern peak in the low resolution 2 cm map of
\citet{1985ApJ...290..108W}. The flux density at 2~cm is about a quarter the peak in the C-configuration maps, so this source is extended. Source C contains about 230 O stars in $0.4\arcsec \times 0.3\arcsec$, based on its 2 cm flux, and 100 equivalent O stars based on the Pa$\,\alpha$ measurement of \citet{2001MNRAS.322..757A}; the discrepancy is probably due to extinction at Pa$\,\alpha$. Source F is an optically thick \HII\ region with spectral index $\alpha > 1.2$. The cluster inferred to be exciting source F has $L_{OB}\sim 1.6 \times 10^{8} L_{\sun}$ and mass $M_{C} < 2.3 \times 10^{5} M_{\sun}$ which is about the mass of a globular cluster.

\subsection{NGC 6946}

NGC 6946 is a barred spiral galaxy close to the Galactic plane ($l=95.72\arcdeg$, $b=11.67\arcdeg$). It is a bright mid-infrared, \citep{1978ApJ...220L..37R}, CO \citep{1985ApJ...298L..21B, 1988PASJ...40..511S, 1988A&A...199...29W,1992A&A...264...55C, 1995ApJ...452L..21R, 1999ApJS..124..403S}, and radio \citep{1983ApJ...268L..79T} source. NGC 6946 contains a moderate starburst of $L_{IR} \sim 2.2 \times 10^{9} L_{\sun}$ and age range 7 - 20~Myr \citep{1996ApJ...467..227E}, and $1 \times 10^{8} M_{\sun}$ of molecular gas, concentrated in the central 175~pc \citep{2004AJ....127.2069M}. The distance to NGC 6946 is estimated at 5.5~Mpc \citep{1988ang..book.....T}, to 5.9 $\pm$ 0.4~Mpc \citep[][based on blue supergiants]{2000A&A...362..544K}; we adopt 5.9~Mpc.

In Figure 5 we show the radio continuum maps and \textit{V}-band and H$\alpha$ \textit{HST} \textit{WFPC2} images of NGC 6946. They were registered by aligning the 6~cm peak and \textit{V}-band and H$\alpha$ peak. We identify five sources with intensity peaks in the emission complex, A - D and I, in the central 60 pc of NGC 6946. The most luminous 2~cm source, Source B, is considered to be with the strongest H$\alpha$ emission. These five sources appear related to giant molecular cloud GMC E1 in \citet{2004AJ....127.2069M}, a cloud that has molecular gas mass of $14 \times 10^{6} M_{\sun}$. Sources E, F and H are further away from the center, and are detected only at 6~cm. They have no optical counterparts although E might be associated with the diffuse optical emission north-west of the optical peak. These nine compact sources contribute 30\,\% of the total 6~cm emission, and 30\,\% of the 2~cm emission, from the central kpc of this galaxy. The high fraction of 6~cm emission from compact sources as compared with the other three spiral galaxies suggests that there are compact and young SNRs in the center of NGC 6946, which may have a more evolved star cluster than the other spirals.

Sources A - D and I are within $2\arcsec$, or 60~pc in projection, of the peak in the \textit{HST} \textit{V}-band image. The infrared peak shown in an archived \textit{NICMOS} \textit{F160W} image lies on source I to within the pointing uncertainties of \textit{HST}. Even with sub-arcsecond resolution, we cannot resolve the extended emission between these sources. We derive their spectral indices based on their peak fluxes. It is possible that they are physically connected to each other and that the two nonthermal sources, A and C, are the consequence of extensive star formation near source B, D, and I. The optical peak of NGC 6946 is associated with a 12 - 15~Myr old Cluster 1094 \citep{2002ApJ...567..896L} of $M_{V}=-10.18$ mag, which is comparable in luminosity to the bright open star clusters but inconspicuous in comparison with a young SSC. By comparison, Source B, D, and I together should have $M_{V} \sim - 12.5$ mag. Using the column density of molecular hydrogen derived from CO \citep{2004AJ....127.2069M}, we estimate that $A_{V} \sim 20 - 30$ mag toward source D and I. Source B, D, and I are probably embedded in GMC E1.

Sources A and C have negative spectra and are likely SNRs \citep{1986ApJ...301..790W}. The 2 cm map shows a decrease in flux density at the position of Source C. This suggests the contamination from extended emission might be responsible for the SNR-like spectral index of these two sources. Thus, we cannot distinguish them from RSNe, SNRs, or \HII\ regions. They could be explained by one or few short-lived young Type I RSN \citep{1998ApJ...500...51W}.

Source E shows extended emission at 2 cm and is marginally detected. In Table 4, we put its peak and integrated fluxes at 2 cm in the area of the 6 cm beam. The flat spectrum suggests that Source E is a large (30 Dor-like) \HII\ region. We do not have significant detections toward Sources F - H at 2 cm, and cannot identify their nature.

\section{Conclusions}

We have imaged compact radio emission in the centers of four nearby spiral galaxies, IC 342, Maffei II, NGC 2903, and NGC 6946, at subarcsecond resolution at 6 and 2 cm with the VLA. Compact radio sources contribute $\sim$ 20 - 30\,\% of the total 2 cm emission from the central kpc of these galaxies, but only 5 - 10\,\% at 6 cm; in NGC 6946, they contribute $\sim$ 30\,\% at both wavelengths. The 2 cm emission within the central 100 - 200~pc of these spirals is dominated by \HII\ regions, while diffuse synchrotron emission dominates at 6 cm. Over half of the 38 identified compact sources appear to be \HII\ regions, based on their radio spectra; of these, one-third to half appear to have optically thick free-free emission. The high resolution of these maps selects for these compact, high brightness \HII\ regions. The brightness temperatures of these optically thick, compact \HII\ regions are consistent with this identification, and indicate actual source diameters of about 1-3~pc ($\sim 0.02\arcsec$ - $0.05\arcsec$.) The compact \HII\ regions contain ionized gas with densities as high as $n_{i} \sim 10^{4}~ cm^{-3}$, which suggests that these \HII\ regions are relatively young. Few of the compact radio sources have candidate counterparts in the visual wavebands. Comparison with optical $HST$ images show that the young \HII\ regions  tend to be found in dust lanes, and probably are subject to high extinction and also dynamical influences from molecular clouds. The predicted infrared emission from these \HII\ regions contributes 3 - 5\,\% of the total infrared luminosity of the galaxies. The largest \HII\ regions require equivalent of $\sim$ 500 - 1,000 O7 stars to excite them; since they are dense, compact \HII\ regions, by analogy with Galactic compact \HII\ regions, they are likely to be very youngest of the massive young clusters, a Myr or less in age.

\acknowledgments

This research has been supported by NSF grants AST 0307950 and AST 0071276. Some of the data presented in this paper were obtained from the Multimission Archive at the Space Telescope Science Institute (MAST). STScI is operated by the Association of Universities for Research in Astronomy, Inc., under NASA contract NAS5-26555. This research has also made use of the NASA/IPAC Extragalactic Database (NED) and the NASA/ IPAC Infrared Science Archive (IRSA) which are operated by the Jet Propulsion Laboratory, California Institute of Technology, under contract with the National Aeronautics and Space Administration.

\clearpage
\begin{figure}
\epsscale{1.0}
\plotone{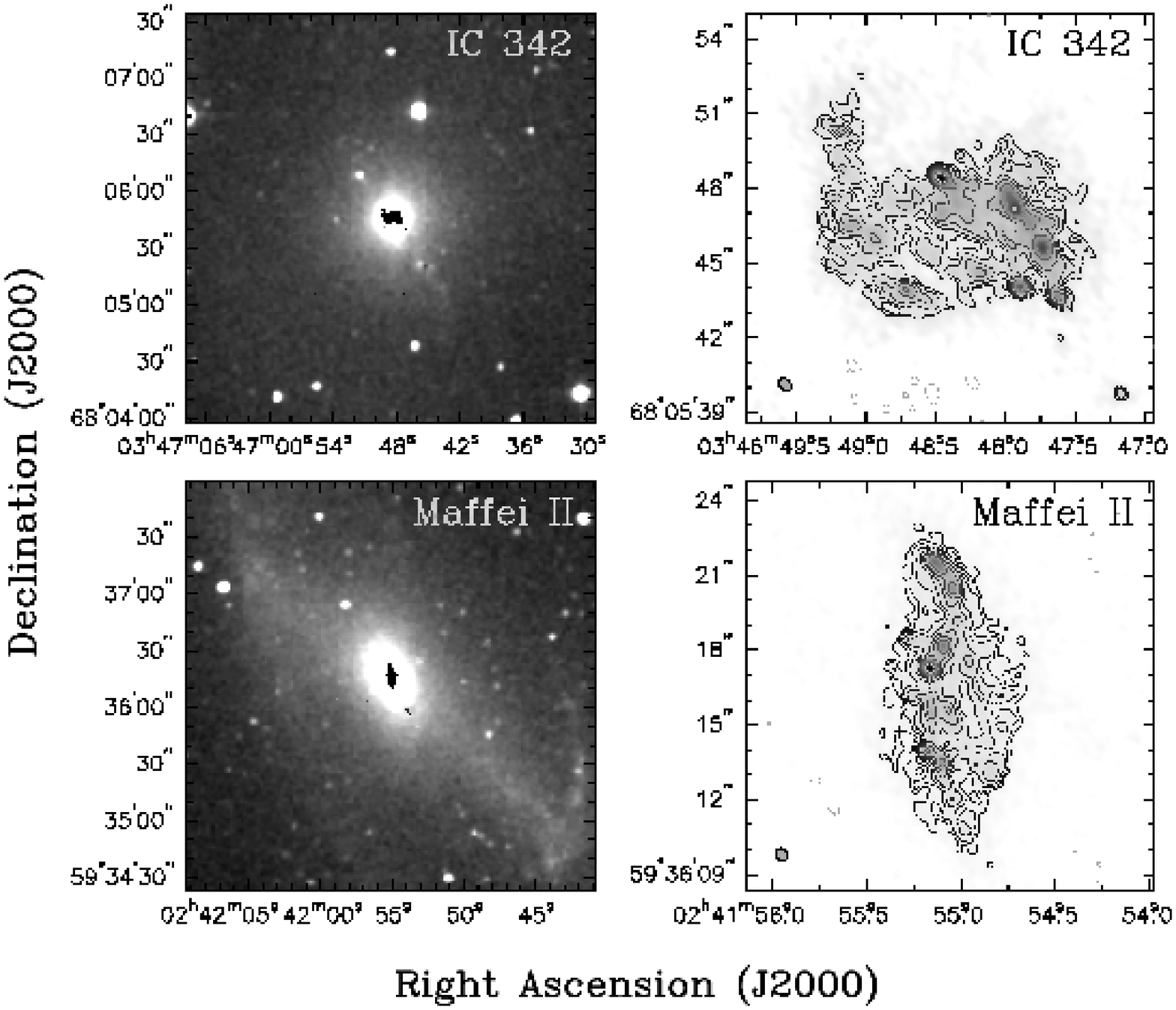}
\caption{\textit{2MASS} \textit{K}-band images and \textit{VLA} 6~cm radio continuum maps of our four sample galaxies. On the left are 3.6$\arcmin$ $\times$ 3.6$\arcmin$ visible images from \textit{2MASS} \textit{K}-band overlaid with 6~cm radio emission from the central region. On the right are 6~cm maps with 16$\arcsec$ $\times$ 16$\arcsec$ field-of-view, contoured at half-integral powers of 2 $\times$ 0.14~mJy beam$^{-1}$ ($\sim$ 4 $\sigma$). Peak flux densities are 1.49, 3.13, 0.37, and 2.74~mJy/beam for IC 342, Maffei II, NGC 2903, and NGC 6946, respectively. Beams are 0\farcs 3 in size.}
\end{figure}

\begin{figure}
\figurenum{1}
\epsscale{1.0}
\plotone{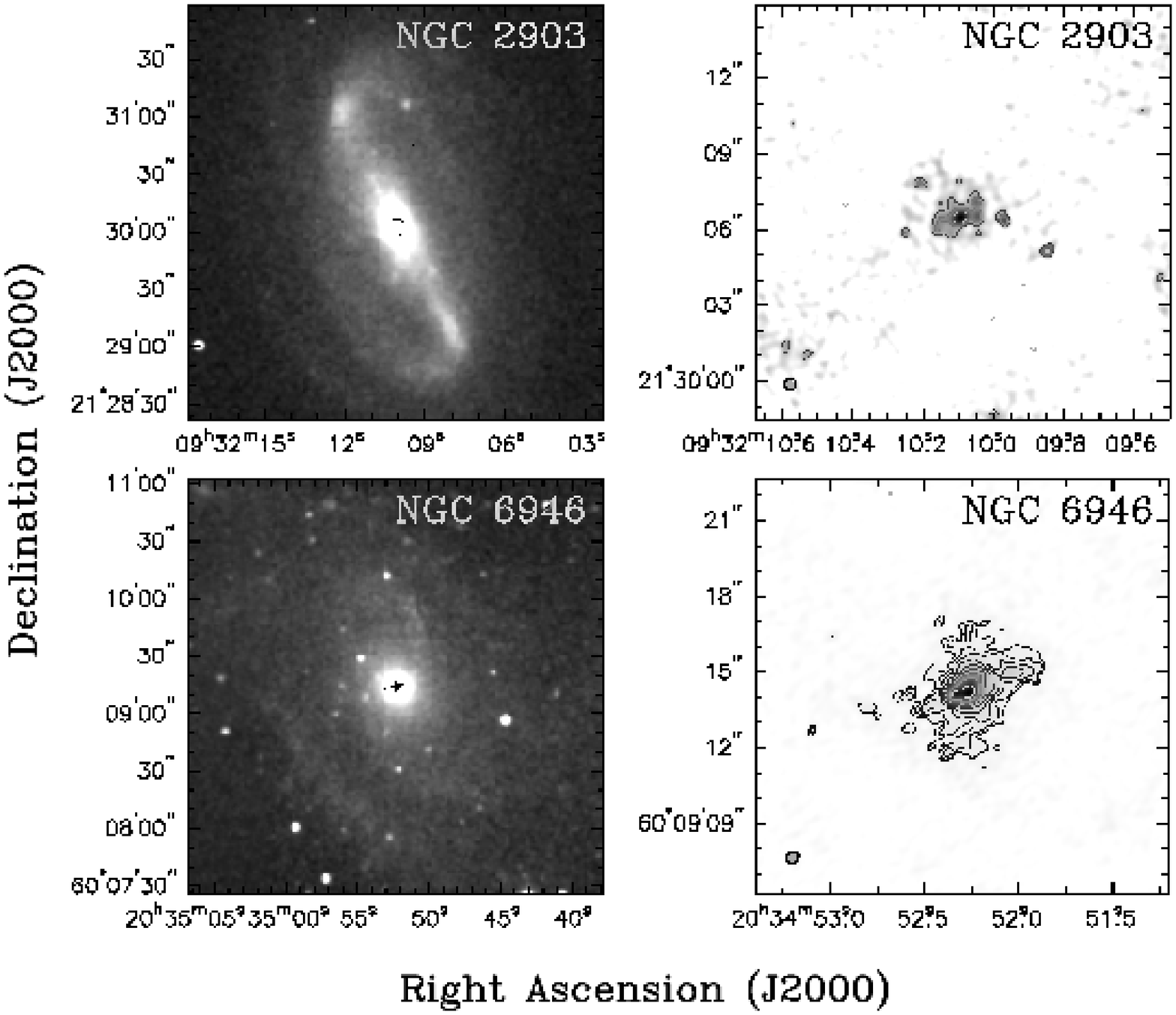}
\caption{Continued.}
\end{figure}

\clearpage
\begin{figure}
\figurenum{2}
\epsscale{0.9}
\plotone{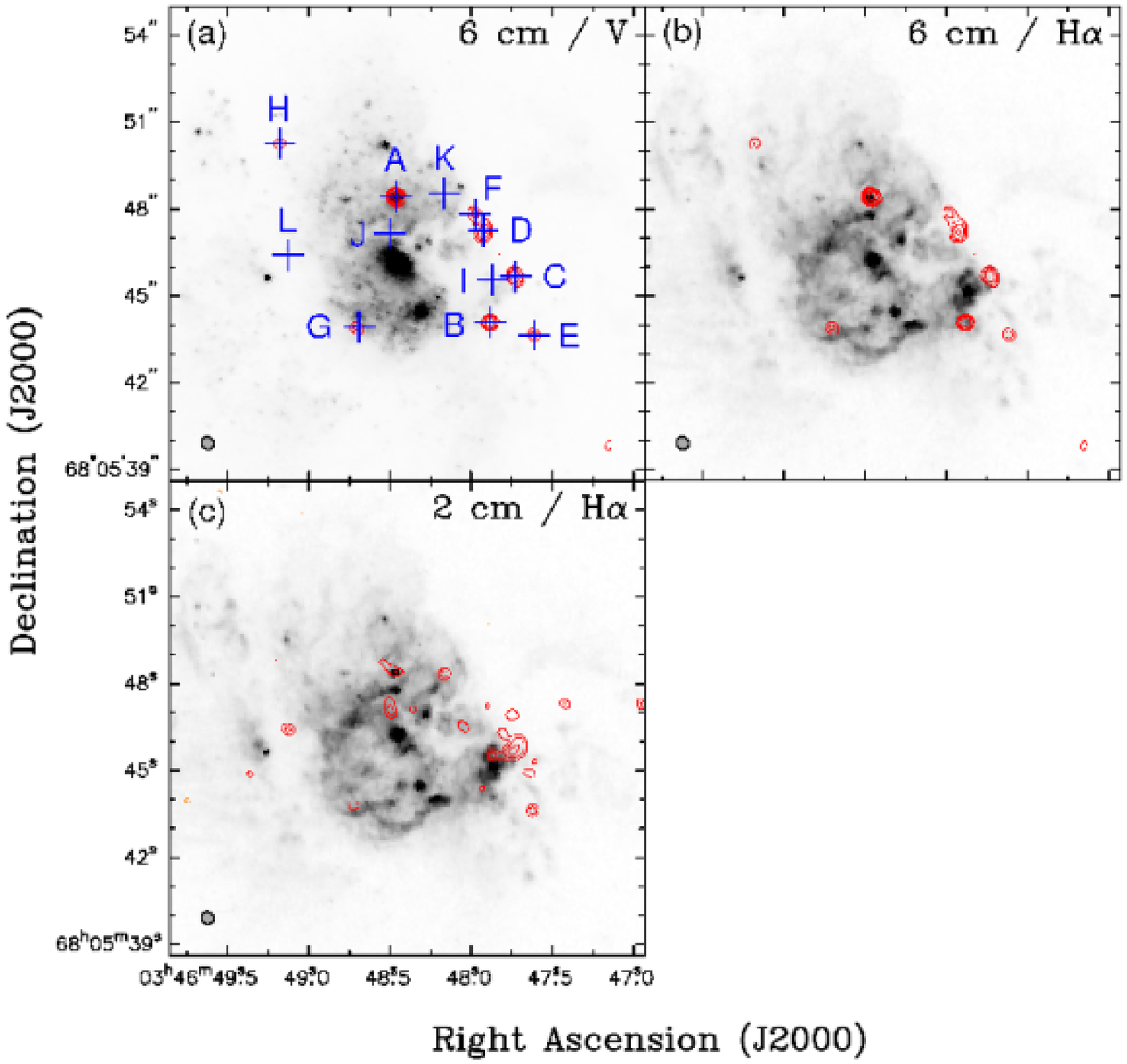}
\caption{Closeups of the central region of IC 342. Upper-left is (a) 6 cm contours overlaid on a \textit{WFPC2} \textit{V}-band (F555W) image. Upper-right, lower-left, and lower-right are (b) 6 cm and (c) 2 cm contours overlaid on a \textit{WFPC2} H$\alpha$ (F656N) image. The beam size of each map has been matched to 0.45$\arcsec$ $\times$ 0.42$\arcsec$, PA$=46\degr$. Contours are half-integral powers of $\pm$ 2 $\times$ 0.30 and 0.43 mJy/beam for 6 cm and 2 cm, respectively. Uncertainty in the registration of the \textit{VLA} and \textit{HST} maps is $\sim 1\arcsec$.}
\end{figure}

\clearpage
\begin{figure}
\figurenum{3}
\epsscale{0.9}
\plotone{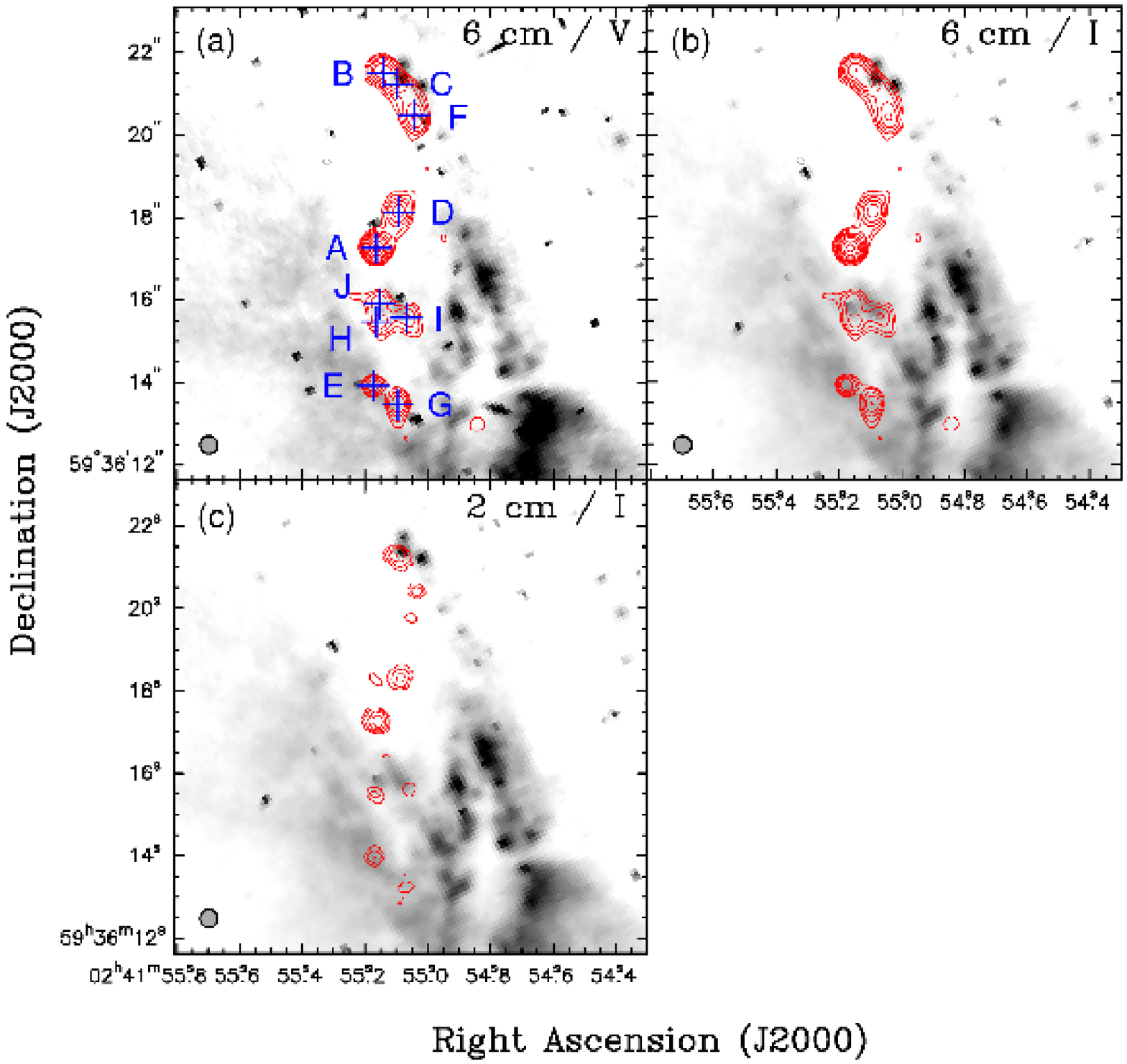}
\caption{Closeups of the central region in Maffei II. Upper-left is (a) 6 cm contours overlaid on a \textit{WFPC2} \textit{V} band (F606W) image. Upper-right and lower-left are (b) 6 cm and (c) 2 cm contours overlaid on a \textit{WFPC2} I band (F814W) image. The beam size of each map has been matched to 0.43$\arcsec$ $\times$ 0.42$\arcsec$, PA$=40\degr$. Contours are at half-integral powers of $\pm$ 2 $\times$ 0.21 and 0.60 mJy/beam for 6 cm and 2 cm maps, respectively. Uncertainty in the registration of the \textit{VLA} and \textit{HST} maps is $\sim 1\arcsec$.}
\end{figure}

\clearpage
\begin{figure}
\figurenum{4}
\epsscale{0.9}
\plotone{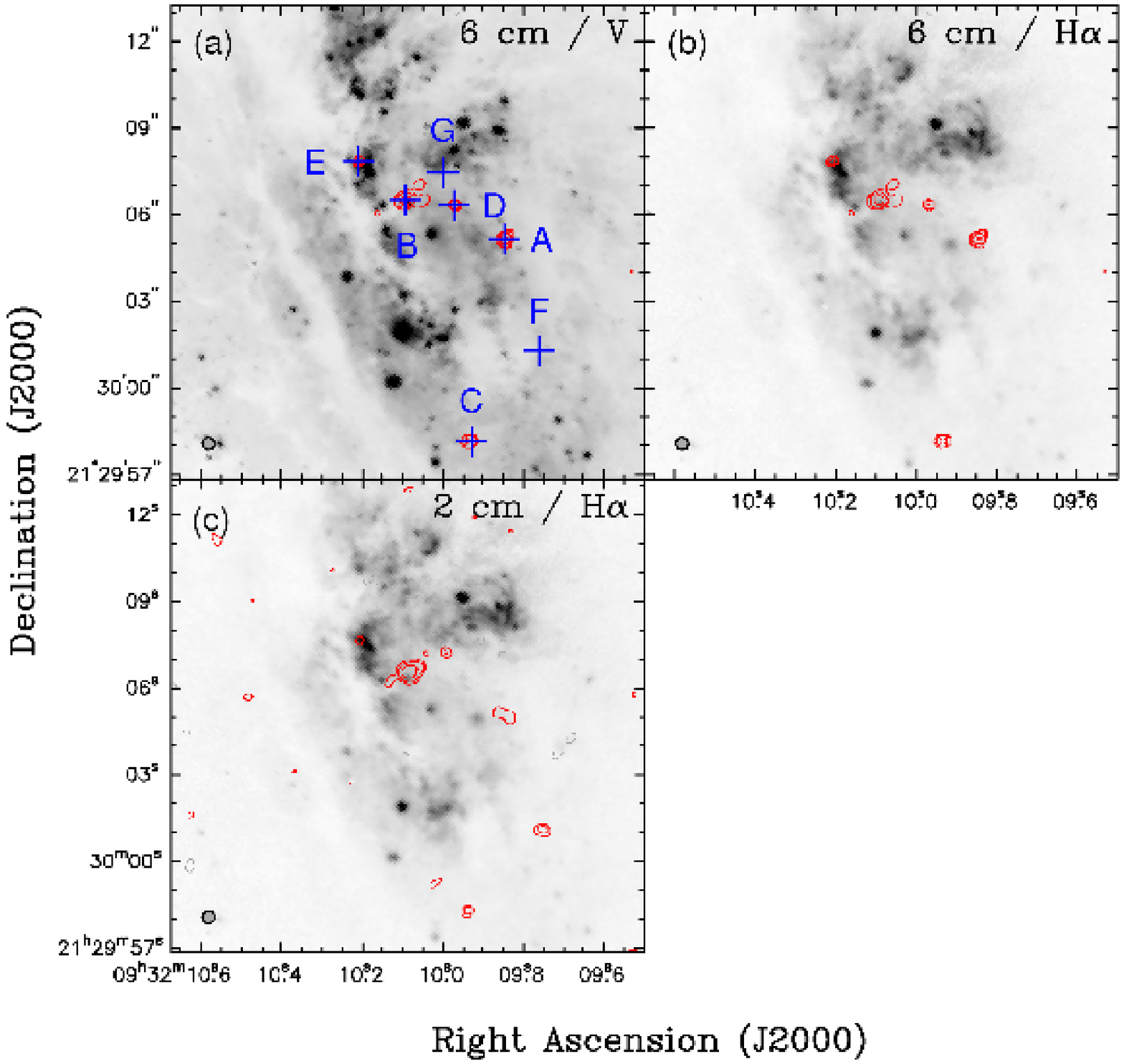}
\caption{Closeups of the central region in NGC 2903. Upper-left is (a) 6 cm contours overlaid on a \textit{WFPC2} \textit{V}-band (F555W) image. Upper-right and lower-left are (b) 6 cm and (c) 2 cm contours overlaid on a \textit{WFPC2} H$\alpha$ (F656N) image. The beam size of each map has been matched to 0.43$\arcsec$ $\times$ 0.41$\arcsec$, PA$=82\degr$. Contours are half-integral powers of $\pm$ 2 $\times$ 0.11 and 0.17 for 6 cm and 2 cm, respectively. Uncertainty in the registration of the \textit{VLA} and \textit{HST} maps is $\sim 1\arcsec$.}
\end{figure}

\clearpage
\begin{figure}
\figurenum{5}
\epsscale{0.9}
\plotone{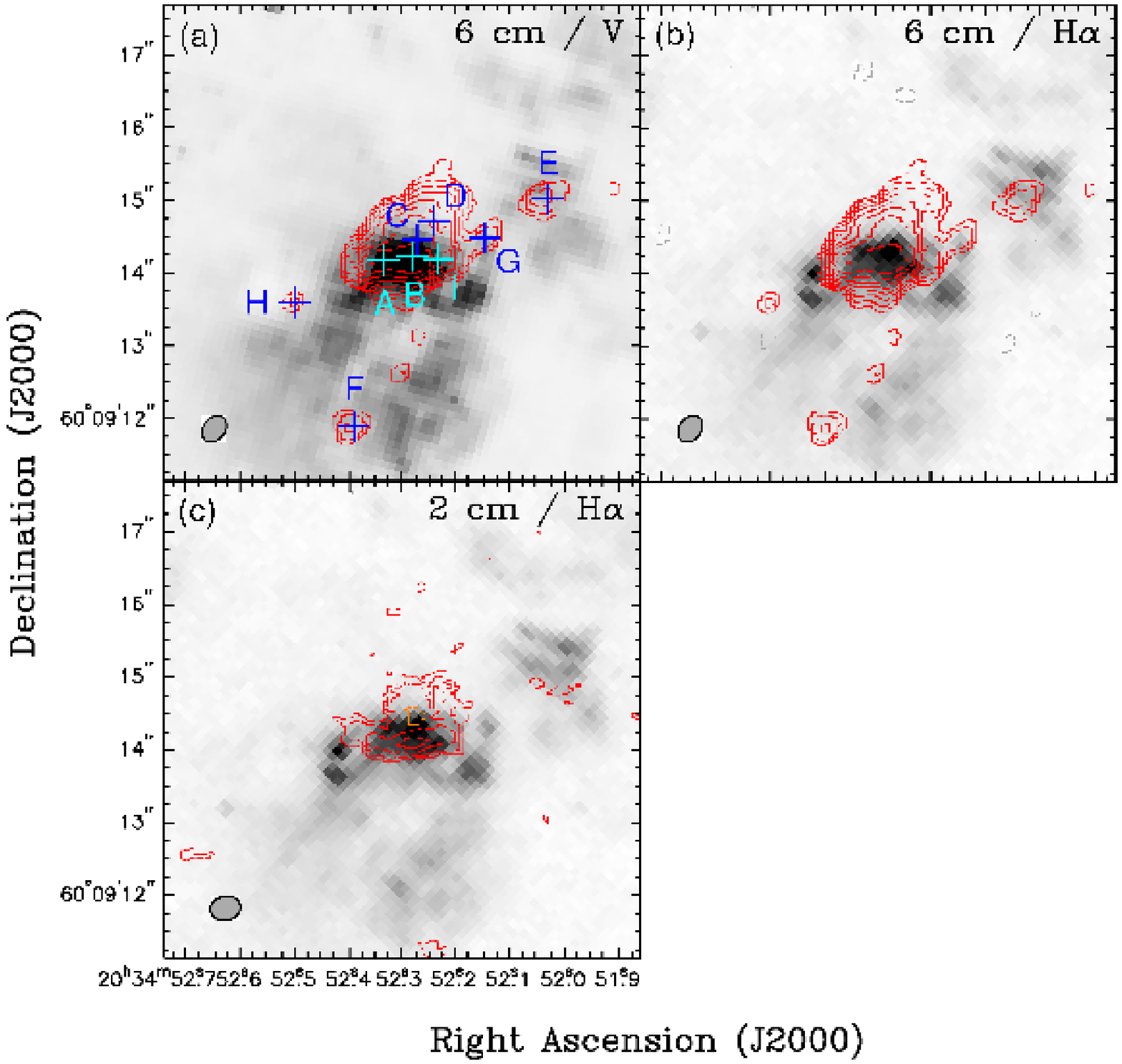}
\caption{Closeups of the center region of NGC 6946. Upper-left is (a) 6 cm contours overlaid on a \textit{WFPC2} \textit{V}-band (F606W) image. Upper-right and the lower-left are (b) 6 cm and (c) 2 cm contours overlaid on \textit{WFPC2} H$\alpha$ (F656N) image. The orange contour represents the region of decreasing flux density toward the center. The beam size of each map has been matched to 0.42$\arcsec$ $\times$ 0.38$\arcsec$, PA$=-79\degr$. Contours are at half-integral powers of $\pm$ 2 $\times$ 0.10 and 0.50 mJy/beam for 6 cm and 2 cm maps, respectively. Uncertainty in the registration of the \textit{VLA} and \textit{HST} maps is $\sim 1\arcsec$.}
\end{figure}

\clearpage
\begin{figure}
\figurenum{6}
\epsscale{0.9}
\plotone{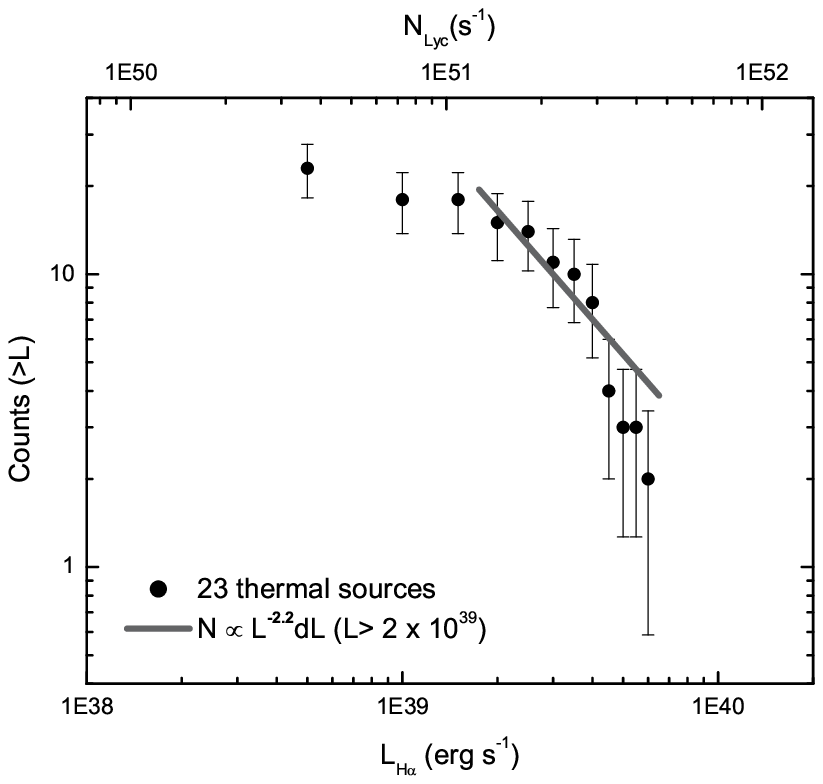}
\caption{Luminosity function of the compact \HII\ regions in the centers of IC 342, Maffei II, NGC 6946, and NGC 2903. Filled circles with statistical error bars show the cumulated number counts of \HII\ regions found in the 4 spiral galaxies of our sample. Solid lines represent the best-fit luminosity function model for sources with $L_{H_{\alpha}} >2 \times 10^{39}$ erg sec$^{-1}$. The  sample is complete down to $N_{Lyc} \sim 6 \times 10^{50}~s^{-1}$ ($H_{\alpha} \sim 8.3 \times 10^{38}~erg~s^{-1}$). Each O7 star contributes roughly $N_{Lyc} \sim 10^{49}~ s^{-1}$; the brightest srouces (in Maffei II and NGC 2903) correspond to clusters of $\sim$ 500 - 1,000 O stars.}
\end{figure}
\clearpage
\begin{deluxetable}{llccccccc}
\tabletypesize{\scriptsize}
\tablewidth{0pt}
\tablecaption{Properties of the Sample Galaxies}
\tablehead{
\multicolumn{1}{l}{Galaxy} &
\multicolumn{1}{l}{Hubble Type} &
\multicolumn{1}{c}{D\tablenotemark{a}} &
\multicolumn{1}{c}{$S_{6}$\tablenotemark{b}} &
\multicolumn{1}{c}{$S_{2}$\tablenotemark{b}} &
\multicolumn{1}{c}{$S_{6}^{map}$\tablenotemark{c}} &
\multicolumn{1}{c}{$S_{2}^{map}$\tablenotemark{c}} &
\multicolumn{1}{c}{$L_{IRAS}$\tablenotemark{d}} &
\multicolumn{1}{c}{$M(H_{2})$\tablenotemark{e}} 
\\
\multicolumn{3}{c}{}&\multicolumn{2}{c}{Total}&\multicolumn{2}{c}{Mapped}&\multicolumn{2}{c}{}\\
\colhead{}
&\colhead{}
&\colhead{(Mpc)}
&\colhead{(mJy)}
&\colhead{(mJy)}
&\colhead{(mJy)}
&\colhead{(mJy)}
&\colhead{$10^{9} L_{\sun}$}
&\colhead{$10^{8} M_{\sun}$}
}

\startdata
IC 342  & SAB(rs)cd & 3.3 & 82 & 38 &15 & 7 & 3.0 & 6.8 \\
Maffei II & SAB(rs)bc & 5.0 & 107 & 46 &16 & 13 & 12.3 & 3.7 \\
NGC 2903 & SB(s)d  & 8.9 & 35 & 12 &5 & 6& 9.4 & 4.5 \\
NGC 6946 & SAB(rs)cd & 5.9 & 39 & 23 &17 &8 & 7.2 & 4.9 \\
\enddata
\tablenotetext{a}{Reference papers for distance: IC 342: \citet{2002AJ....124..839S}; Maffei II: \citet{1993AJ....105..121H}; NGC 2903: \citet{2000A&AS..142..425D}; NGC 6946: \citet{2000A&A...362..544K}.}
\tablenotetext{b}{Total radio continuum fluxes in \textit{VLA} low resolution maps using B-config. at 6 cm and C-config at 2 cm for IC 342, NGC 6946 \citep{1983ApJ...268L..79T}, and Maffei II \citep{1994ApJ...421..122T}; using C-config. at 6 cm for NGC 2903 \citep{1985ApJ...290..108W}.}
\tablenotetext{c}{Total fluxes in naturally-weighted, 0\farcs3 maps (this paper).}
\tablenotetext{d}{$L_{IRAS}$ (infrared luminosity) calculated using the prescription from \citet{1996ARA&A..34..749S}. \textit{IRAS} data are from point source flux densities listed in the \textit{IRAS Faint Source Catalog}, version 2.0 \citep{1990IRASF.C......0M}. \textit{IRAS} fluxes of Maffei II are from \citet{1986A&A...166L...8C}.}
\tablenotetext{e}{Molecular mass calculated using $^{12}CO$ data \citep{1990ApJ...351..422S} with distance correction.}
\end{deluxetable}

\clearpage
\begin{deluxetable}{llrclll}
\tabletypesize{\scriptsize}
\tablewidth{0pt} \tablecaption{\textit{VLA} Data} 
\tablehead{
\multicolumn{1}{l}{Object/Band\tablenotemark{a}} &
\multicolumn{1}{l}{Obs. Date} &
\multicolumn{1}{c}{On Source} &
\multicolumn{1}{c}{Config.} &
\multicolumn{1}{l}{Flux Calib.} &
\multicolumn{1}{l}{Phase Calib.} &
\multicolumn{1}{c}{Prog. ID}
\\
\colhead{}
&\colhead{mm/dd/yyyy}
&\colhead{(sec)}
&\colhead{}
&\colhead{}
&\colhead{}
&\colhead{}
}
\startdata
IC 342\\
\dots \dots C-band 
    & 11/23/1996 & 5290 & A & 3C48      & 0224+671 & AW419 \\
    & 11/20/1988 & 4040 & A & 3C48 \& 3C147  & 0224+671 & AT98\tablenotemark{b}\\
    & 07/11/1981 & 6260 & A & 3C286      & 0212+735 & AT21 \\
\dots \dots U-band 
    & 08/23/1999 & 4310 & A & 3C286      & 0228+673 & AT227\tablenotemark{b}\\
    & 12/08/1995 & 4550 & B & 3C48      & 0224+671 & AW419 \\
    & 11/14/1981 & 6020 & C & 3C286      & 0224+671 & AT21 \\
Maffei II\\
\dots \dots C-band 
    & 08/26/1989 & 6860 & BC & 3C138 \& 3C48  & 0224+671 & AH353 \\
    & 11/20/1988 & 4020 & A & 3C48 \& 3C147  & 0224+671 & AT98\tablenotemark{b}\\
    & 09/05/1982 & 2970 & B & 3C286      & 0224+671 & AT21 \\
    & 09/04/1982 & 5570 & B & 3C286      & 0224+671 & AT21 \\
\dots \dots U-band 
    & 10/12/1983 & 2330 & A & 3C48 \& 3C138  & 0224+671 & AH141\tablenotemark{b}\\
    & 04/18/1983 & 4090 & C & 3C286      & 0224+671 & AT25 \\
    & 04/18/1983 & 1170 & C & 3C286      & 0224+671 & AT25 \\
NGC 2903\\
\dots \dots C-band 
    & 01/11/1997 & 4340 & A & 3C286     & 0851+202 & AW419 \\
    & 11/26/1996 & 4320 & A & 3C286     & 0851+202 & AW419 \\
    & 11/18/1988 & 3470 & A & 3C286     & 0851+202 & AT98\tablenotemark{b}\\
    & 02/01/1988 & 870  & B & 3C286     & 0953+254 & AS314 \\
    & 08/16/1986 & 870  & B & 3C48      & 0953+254 & AD176 \\
    & 04/26/1985 & 5530 & B & 3C286     & 0851+202 & AB324 \\
    & 10/24/1982 & 8550 & B & 3C286     & 0851+202 & AT21 \\
\dots \dots U-band 
    & 08/23/1999 & 4280 & A & 3C286     & 0854+201 & AT227\tablenotemark{b}\\
    & 04/14/1996 & 1500 & C & 3C286 \& 3C147 & 1004+141 & AQ11 \\
    & 01/05/1996 & 3840 & B & 3C48      & 0851+202 & AW419 \\
    & 12/27/1995 & 1800 & B & 3C286     & 0851+202 & AW419 \\
    & 12/08/1986 & 17100 & C & 3C286     & 0851+202 & AB324 \\
    & 09/18/1985 & 9480 & C & 3C286     & 0851+202 & AB324 \\
    & 04/18/1983 & 2210 & C & 3C286     & 0851+202 & AT25 \\
    & 04/17/1983 & 2900 & C & 3C286     & 0851+202 & AT25 \\
NGC 6946\\
\dots \dots C-band 
    & 06/17/1999 & 3680 & AD & 3C48      & 2021+614 & AS568 \\
    & 11/23/1996 & 4730 & A & 3C48      & 2021+614 & AW419 \\
    & 04/10/1994 & 6340 & A & 3C286     & 2021+614 & AS525 \\
    & 11/20/1988 & 4010 & A & 3C48 \& 3C147 & 2021+614 & AT98\tablenotemark{b}\\
\dots \dots U-band 
    & 12/08/1995 & 4580 & B & 3C48      & 2021+614 & AW419 \\
    & 10/12/1983 & 2600 & A & 3C48 \& 3C138 & 2021+614 & AH141\tablenotemark{b}\\
    & 11/14/1981 & 6020 & C & 3C286     & 2021+614 & AT21 \\
\enddata
\tablenotetext{a}{\textit{VLA} \textit{C}-band: 6 cm; \textit{U}-band: 2 cm; \textit{K}-band: 1.3 cm.}
\tablenotetext{b}{Unpublished data from observations executed by the authors.}
\end{deluxetable}

\clearpage
\begin{deluxetable}{rccclclclclc}
\tabletypesize{\scriptsize}
\tablewidth{0in}
\tablecaption{Compact Continuum Source Observed Quantities}
\tablehead{
\multicolumn{1}{l}{Source} &
\multicolumn{1}{c}{$\alpha$(J2000)} &
\multicolumn{1}{c}{$\delta$(J2000)} &
\multicolumn{1}{c}{$S_{6 cm}^{Peak}$\tablenotemark{;a,c}} &
\multicolumn{1}{c}{$S_{6 cm}$\tablenotemark{b,c}} &
\multicolumn{1}{c}{$S_{2 cm}^{Peak}$\tablenotemark{;a,c}} &
\multicolumn{1}{c}{$S_{2 cm}$\tablenotemark{b,c}} &
\multicolumn{1}{c}{Size\tablenotemark{d}}\\
\colhead{} &
\colhead{($^{s}$)} &
\colhead{($\arcsec$)} &
\colhead{(mJy beam$^{-1})$} &
\colhead{(mJy)} &
\colhead{(mJy beam$^{-1})$} &
\colhead{(mJy)} &
\colhead{$\arcsec$ $\times$ $\arcsec$, $\degr$}}
\startdata
IC 342  ....... & 03$^{h}$46$^{m}$+ & 68$\arcdeg$05$\arcmin$+\\
A & 48.46 & 48.4 & 1.4   & 1.5 $\pm$ 0.2 & 0.6   & 0.9 $\pm$ 0.3          & 0.2 0.1, 40 \\
B & 47.88 & 44.1 & 0.8   & 0.8 $\pm$ 0.2 & $<$ 0.3\tablenotemark{e} & 0.4 $\pm$ 0.3  & 0.3 0.1, 120\\
C & 47.73 & 45.7 & 0.8   & 1.3 $\pm$ 0.3 & 1.0   & 2.2 $\pm$ 0.5          & 0.4 0.3, 20 \\
D & 47.93 & 47.3 & 0.8   & 1.5 $\pm$ 0.4 & $<$ 0.3\tablenotemark{e} & $<$ 0.3     & 0.6 0.4, 10 \\
E & 47.61 & 43.7 & 0.5   & 0.7 $\pm$ 0.3 & 0.7   & 0.7 $\pm$ 0.5          & 0.4 0.3, 40 \\
F & 47.98 & 47.8 & 0.5   & 0.6 $\pm$ 0.3 & $<$ 0.3\tablenotemark{e} & $<$ 0.3     & 0.4 0.3, 70 \\
G & 48.69 & 44.0 & 0.5   & 0.6 $\pm$ 0.3 & 0.5   & 0.5 $\pm$ 0.5          & 0.5 0.3, 100\\
H & 49.18 & 50.3 & 0.4   & 0.5 $\pm$ 0.3 & $<$ 0.3\tablenotemark{e} & $<$ 0.3     & 0.4 0.3, 140 \\
I & 47.86 & 45.5 & $<$ 0.2\tablenotemark{e} & $<$ 0.2     & 0.7   & 0.7 $\pm$ 0.5 & 0.4 0.3, 120\\
J & 48.50 & 47.2 & $<$ 0.2\tablenotemark{e} & $<$ 0.2     & 0.7   & 0.7 $\pm$ 0.5 & 0.8 0.2, 10 \\
K & 48.17 & 48.5 & $<$ 0.2\tablenotemark{e} & $<$ 0.2     & 0.7   & 0.7 $\pm$ 0.5 & 0.4 0.3, 150\\
L & 48.99 & 46.4 & $<$ 0.2\tablenotemark{e} & $<$ 0.2     & 0.7   & 0.7 $\pm$ 0.4 & 0.4 0.2, 70\\
\\
Maffei II ....... & 02$^{h}$41$^{m}$+ & 59$\arcdeg$36$\arcmin$+\\
A & 55.16 & 17.3 & 2.6   & 3.1 $\pm$ 0.2 & 2.1   & 2.7 $\pm$ 0.5           & 0.3 0.2, 60 \\
B & 55.14 & 21.5 & 1.1   & 1.1 $\pm$ 0.2 & $<$ 0.3\tablenotemark{e} & $<$ 0.3     & 0.4 0.3, 60 \\
C & 55.10 & 21.2 & 1.0   & 1.6 $\pm$ 0.2 & 1.7   & 2.2 $\pm$ 0.6           & 0.4 0.3, 70 \\
D & 55.09 & 18.1 & 1.0   & 1.5 $\pm$ 0.3 & 1.3   & 1.8 $\pm$ 0.7           & 0.5 0.4, 150\\
E & 55.17 & 13.9 & 0.8   & 0.8 $\pm$ 0.2 & 1.2   & 1.3 $\pm$ 0.5           & 0.3 0.2, 160\\
F & 55.04 & 20.5 & 0.8   & 1.6 $\pm$ 0.2 & 1.0   & 1.0 $\pm$ 0.4           & 0.2 0.1, 70 \\
G & 55.10 & 13.5 & 0.8   & 1.2 $\pm$ 0.2 & 0.8   & 1.5 $\pm$ 0.4           & 0.5 0.1, 0 \\
H & 55.16 & 15.5 & 0.7   & 1.0 $\pm$ 0.3 & 0.8   & 1.4 $\pm$ 0.7           & 0.5 0.4, 10 \\
I & 55.07 & 15.6 & 0.6   & 0.9 $\pm$ 0.3 & 0.8   & 0.8 $\pm$ 0.8           & 0.6 0.4, 40 \\
J & 55.15 & 15.9 & 0.6   & 0.7 $\pm$ 0.3 & $<$ 0.3\tablenotemark{e} & $<$ 0.3     & 0.5 0.4, 80 \\
\\
NGC 2903 ....... & 09$^{h}$32$^{m}$+ & 21$\arcdeg$30$\arcmin$+\\
A & 09.85 & 05.2 & 0.3   & 0.4 $\pm$ 0.1 & 0.2   & 0.4 $\pm$ 0.2          & 0.4 0.2, 150\\
B & 10.10 & 06.5 & 0.3   & 0.5 $\pm$ 0.2 & 0.4   & 1.2 $\pm$ 0.4          & 0.6 0.4, 120\\
C & 09.93 & -01.9 & 0.3   & 0.4 $\pm$ 0.2 & 0.3   & 0.3 $\pm$ 0.3          & 0.4 0.3, 100\\
D & 09.97 & 06.3 & 0.2   & 0.2 $\pm$ 0.1 & $<$ 0.1\tablenotemark{e} & $<$ 0.1     & 0.2 0.2, 60 \\
E & 10.21 & 07.9 & 0.2   & 0.2 $\pm$ 0.1 & 0.2   & 0.4 $\pm$ 0.2          & 0.4 0.2, 140\\
F & 09.76 & 01.3 & $<$ 0.1\tablenotemark{e} & $<$ 0.1    & 0.3   & 0.4 $\pm$ 0.3 & 0.5 0.4, 90 \\
G & 10.00 & 07.5 & $<$ 0.1\tablenotemark{e} & $<$ 0.1    & 0.2   & 0.3 $\pm$ 0.2 & 0.4 0.2, 50 \\
\\
NGC 6946 ....... & 20$^{h}$34$^{m}$+ & 60$\arcdeg$09$\arcmin$+\\
A & 52.34 & 14.2 & 1.5 &   & 1.1   &  & confused\\
B & 52.28 & 14.2 & 1.4 &   & 1.5   &  & confused\\
C & 52.28 & 14.5 & 1.1 &   & 0.7\tablenotemark{f} &  & confused\\
D & 52.25 & 14.7 & 1.0\tablenotemark{f} &   & 1.0 &  & confused\\
E & 52.03 & 15.0 & 0.3 & 0.5 $\pm$ 0.2 & 0.5   & 0.5 $\pm$ 0.5                  & 0.5 0.3, 120\\
F & 52.39 & 11.9 & 0.2 & 0.3 $\pm$ 0.1 & $<$ 0.4\tablenotemark{e} & $<$ 0.4            & 0.3 0.2, 70 \\
G & 52.15 & 14.5 & 0.2 & 0.3 $\pm$ 0.3 & $<$ 0.4\tablenotemark{e} & $<$ 0.4            & 0.9 0.5, 130\\
H & 52.50 & 13.6 & 0.1 & 0.3 $\pm$ 0.1 & $<$ 0.4\tablenotemark{e} & $<$ 0.4            & 0.4 0.2, 140\\
I & 52.23 & 14.2 & 0.4\tablenotemark{f} &   & 1.3 &   & confused\\
\enddata
\tablenotetext{a}{$S_{6 cm}^{Peak}$ and $S_{2 cm}^{Peak}$: peak flux density (intensity) of 6~cm and 2~cm continuum, respectively, with beams 0.45$\arcsec$ $\times$ 0.42$\arcsec$, PA = 46$\degr$ for IC 342, 0.43$\arcsec$ $\times$ 0.42$\arcsec$, PA = 40$\degr$ for Maffei II, 0.43$\arcsec$ $\times$ 0.42$\arcsec$, PA = 82$\degr$ for NGC 2903, and 0.42$\arcsec$ $\times$ 0.38$\arcsec$, PA = $-79\degr$ for NGC 6949. The rms noise levels at 6~cm are 0.05, 0.04, 0.03, and 0.03 mJy/beam for IC 342, Maffei 2, NGC 2903, and NGC 6946, respectively, and at  2~cm are 0.09, 0.11, 0.05, and 0.14 mJy/beam.}
\tablenotetext{b}{$S_{6 cm}$ and $S_{2 cm}$: total integrated fluxes  at 6~cm and 2~cm, respectively. Quoted uncertainties in the fluxes are derived from the rms and measured source size; intensities are more accurate for extended sources. The systematic error of flux is expected to be $\lesssim$ 5\,\%.}
\tablenotetext{c}{Flux densities are measured in (\textit{u,v})-restricted map. The uncertainties quoted here are 3~$\sigma$ See text for details.}
\tablenotetext{d}{Single Gaussian fitting uncertainty is $\lesssim$ 30\,\%, using \textit{AIPS} task \textit{IMFIT}.}
\tablenotetext{e}{3 $\sigma$ upper limit.}
\tablenotetext{f}{Peak flux density is averaged flux density over a beam area.}
\end{deluxetable}

\clearpage
\begin{deluxetable}{rcccccccccc}
\tabletypesize{\scriptsize}
\tablewidth{0in}
\tablecaption{Compact Continuum Source Derived Properties}
\tablehead{
\multicolumn{1}{l}{Source} &
\multicolumn{1}{c}{$\alpha_{6-2}$\tablenotemark{a}} &
\multicolumn{1}{c}{Type} &
\multicolumn{1}{c}{Free-Free} &
\multicolumn{1}{c}{$\rm N_{Lyc}^{thin}$\tablenotemark{;b}} &
\multicolumn{1}{c}{$\rm N_{O7}$\tablenotemark{c}} &
\multicolumn{1}{c}{$\rm L_{IR}$\tablenotemark{d}} &
\multicolumn{1}{c}{Comments} \\
\colhead{} &
\colhead{} &
\colhead{} &
\colhead{$\tau$} &
\colhead{$10^{51}$ $s^{-1}$} &
\colhead{} &
\colhead{$10^{7} \rm L_{\sun}$} &
\colhead{}}
\startdata
IC 342 \dots \\
A  & -0.5 $^{+0.2}_{-0.2}$ & SNR   & N/A  &   &   &   & \\ 
B & -0.6 $^{+0.4}_{-0.5}$ & SNR   & N/A  &   &   &   & \\ 
C  & 0.5 $^{+0.2}_{-0.2}$ & \HII\  & Thick & 2.1 & 210 & 4.2 & \\ 
D  & $<$ -1.5       & SNR/RSN   & N/A  &   &   &   & \\ 
E  & 0.0 $^{+0.5}_{-0.5}$ & \HII\ ? & Thin? & 0.7 & 70 & 1.4 & \\ 
F  & $<$ -0.6       & SNR   & N/A  &   &   &   & \\ 
G  & -0.3 $^{+0.6}_{-0.8}$ & SNR   & N/A  &   &   &   & \\ 
H  & $<$ -0.5       & SNR   & N/A  &   &   &   & \\ 
I  & $>$ 1.1        & \HII\  & Thick & 0.7 & 70 & 1.4 & \\
J  & $>$ 1.1        & \HII\  & Thick & 0.7 & 70 & 1.4 & \\ 
K  & $>$ 1.1        & \HII\  & Thick & 0.7 & 70 & 1.4 & \\
L  & $>$ 1.1        & \HII\  & Thick & 0.7 & 70 & 1.4 & \\
Total &      &            &     & 5.6 & 560 & 11.1 & \\
Maffei II \dots \\
A  & -0.1 $^{+0.1}_{-0.1}$ & \HII\  & Thin & 6.1 & 610 & 12.2 & \\ 
B  & $<$ -1.2        & SNR/RSN   & N/A  &    &   &   & \\
C  &  0.3 $^{+0.2}_{-0.2}$ & \HII\  & Thick & 4.9 & 490 & 9.7 & \\ 
D  &  0.2 $^{+0.2}_{-0.3}$ & \HII\  & Thin? & 4.1 & 410 & 8.2 & \\ 
E  &  0.4 $^{+0.3}_{-0.3}$ & \HII\  & Thick? & 2.9 & 290 & 5.8 & \\ 
F  & -0.4 $^{+0.2}_{-0.2}$ & SNR   & N/A  &    &   &   & \\ 
G  &  0.2 $^{+0.2}_{-0.2}$ & \HII\  & Thin? & 3.4 & 340 & 6.8 & \\ 
H  &  0.3 $^{+0.3}_{-0.4}$ & \HII\  & Thick? & 3.2 & 320 & 6.5 & \\ 
I  & -0.1 $^{+0.5}_{-0.7}$ & \HII\ ? & Thin? & 1.9 & 190 & 3.7 & \\ 
J  & $<$ -0.7        & SNR   & N/A  &    &   &   & \\ 
Total &      &            &     & 26 & 2600 & 53 & \\
NGC 2903 \dots \\
A  &  0.0 $^{+0.3}_{-0.3}$ & \HII\ ? & Thin? & 2.9 & 290 & 5.8 & \\ 
B  &  0.8 $^{+0.3}_{-0.3}$ & \HII\  & Thick & 8.2 & 820 & 16 & \\ 
C  & -0.2 $^{+0.3}_{-0.4}$ & \HII\  & Thin? & 2.3 & 230 & 4.6 & \\ 
D  & $<$ -0.5        & SNR   & N/A  &    &   &   & \\ 
E  &  0.6 $^{+0.4}_{-0.4}$ & \HII\  & Thick & 2.9 & 290 & 5.8 & \\ 
F  & $>$ 1.2        & \HII\  & Thick & 3.1 & 310 & 6.1 & \\
G  & $>$ 1.0        & \HII\  & Thick & 2.1 & 210 & 4.2 & \\ 
Total &      &            &     & 21 & 2100 & 43 & \\
NGC 6946 \dots \\
A\tablenotemark{e}  & -0.3 $^{+0.2}_{-0.2}$ & SNR?  & N/A  &    &   &   & \\
B\tablenotemark{e}  & 0.1 $^{+0.1}_{-0.2}$ & \HII\ ?  &  Thin?  & 4.5 & 450 & 9.0 & \\
C\tablenotemark{e}  & $<$ -0.4        & SNR?  & N/A  &     &     &   & hole at 2cm \\
D\tablenotemark{e}  &  0.0 $^{+0.2}_{-0.2}$ & \HII\ ? & Thin?   & 3.0 & 300 & 6.0 & \\
E  &  0.0 $^{+0.4}_{-0.4}$ & \HII\ ? & Thin? & 1.5 & 150 & 2.7 & \\
F  & $<$ 0.3        & ?   &    &    &   &   & undetermined \\
G  & $<$ 0.3        & ?   &    &    &   &   & undetermined \\
H  & $<$ 0.3        & ?   &    &    &   &   & undetermined \\
I\tablenotemark{e} & $>$ 1.1        & \HII\ & Thick & 4.5 & 450 & 9.1 & \\
Total &      &            &    & 14 & 1400 & 27 & \\
\enddata
\tablenotetext{a}{Spectral indices are computed from the total flux densities. The upper and lower limits are based on $\sqrt{2}$~$\sigma$ release at both wavelengths.}
\tablenotetext{b}{Assume the electron temperature $T_{e} =$ 10,000~K. The values will be $\sim$ 15\,\% lower than listed if $T_{e} =$ 6,000~K is assumed.}
\tablenotetext{c}{Assume the ionization rate of a reference O star, $N_{O7} \approx N_{Lyc} / 10^{49} sec^{-1}$.}
\tablenotetext{d}{Using $\rm L_{IR}/N_{Lyc} = 2 \times 10^{-44} (L_{\sun} \, sec)$ for $L_{IR}$ estimate, based on ZAMS cluster (following \citealt{1998AJ....116.1212T}), and assuming $L_{IR} = L_{OB}$.}
\tablenotetext{e}{confused sources. $\alpha$ are derived from peak fluxes. The $N_{Lyc}^{thin}$, $N_{O7}$, and $L_{IR}$ of them might be overestimated due to possible contamination in peak flux measurements by extended emission.}
\end{deluxetable}
\end{document}